\documentstyle[12pt]{article}

\catcode`\@=11
\long\def\@makefntext#1{
\protect\noindent \hbox to 3.2pt {\hskip-.9pt
$^{{\ninerm\@thefnmark}}$\hfil}#1\hfill}		

\def\@makefnmark{\hbox to 0pt{$^{\@thefnmark}$\hss}}  

\def\ps@myheadings{\let\@mkboth\@gobbletwo
\def\@oddhead{\hbox{}
\rightmark\hfil\ninerm\thepage}
\def\@oddfoot{}\def\@evenhead{\ninerm\thepage\hfil
\leftmark\hbox{}}\def\@evenfoot{}
\def\sectionmark##1{}\def\subsectionmark##1{}}
\renewcommand{\thefootnote}{\fnsymbol{footnote}}

\newcounter{sectionc}\newcounter{subsectionc}\newcounter{subsubsectionc}
\renewcommand{\section}[1] {\vspace*{0.6cm}\addtocounter{sectionc}{1}
\setcounter{subsectionc}{0}\setcounter{subsubsectionc}{0}\noindent
	{\normalsize\bf\thesectionc. #1}\par\vspace*{0.4cm}}
\renewcommand{\subsection}[1] {\vspace*{0.6cm}\addtocounter{subsectionc}{1}
	\setcounter{subsubsectionc}{0}\noindent
    {\normalsize\it\thesectionc.\thesubsectionc. #1}\par\vspace*{0.4cm}}
\renewcommand{\subsubsection}[1]
{\vspace*{0.6cm}\addtocounter{subsubsectionc}{1}
\noindent {\normalsize\rm\thesectionc.\thesubsectionc.\thesubsubsectionc.
    #1}\par\vspace*{0.4cm}}

\newcounter{appendixc}
\newcounter{subappendixc}[appendixc]
\newcounter{subsubappendixc}[subappendixc]

\renewcommand{\appendix}[1] {\vspace*{0.6cm}
        \refstepcounter{appendixc}
        \setcounter{figure}{0}
        \setcounter{table}{0}
        \setcounter{equation}{0}
        \renewcommand{\thefigure}{\Alph{appendixc}.\arabic{figure}}
        \renewcommand{\thetable}{\Alph{appendixc}.\arabic{table}}
        \renewcommand{\theappendixc}{\Alph{appendixc}}
        \renewcommand{\theequation}{\Alph{appendixc}.\arabic{equation}}
        \noindent{\bf Appendix \theappendixc #1}\par\vspace*{0.4cm}}

\renewenvironment{thebibliography}[1]
    {\begin{list}{\arabic{enumi}.}
    {\usecounter{enumi}\setlength{\parsep}{0pt}
\setlength{\leftmargin 1.25cm}{\rightmargin 0pt}
     \setlength{\itemsep}{0pt} \settowidth
    {\labelwidth}{#1.}\sloppy}}{\end{list}}

\newcounter{itemlistc}
\newcounter{romanlistc}
\newcounter{alphlistc}
\newcounter{arabiclistc}

\newcommand{\fcaption}[1]{
        \refstepcounter{figure}
        \setbox\@tempboxa = \hbox{\footnotesize Fig.~\thefigure. #1}
        \ifdim \wd\@tempboxa > 6in
           {\begin{center}
        \parbox{6in}{\footnotesize\baselineskip=15pt Fig.~\thefigure. #1}
            \end{center}}
        \else
             {\begin{center}
             {\footnotesize Fig.~\thefigure. #1}
              \end{center}}
        \fi}
\newcommand{\tcaption}[1]{
        \refstepcounter{table}
        \setbox\@tempboxa = \hbox{\footnotesize Table~\thetable. #1}
        \ifdim \wd\@tempboxa > 6in
           {\begin{center}
        \parbox{6in}{\footnotesize\baselineskip=15pt Table~\thetable. #1}
            \end{center}}
        \else
             {\begin{center}
             {\footnotesize Table~\thetable. #1}
              \end{center}}
        \fi}

 1 
 1  
 1  
 1  
 1  
 1  
 1  

\font\ninerm=cmr9

\evensidemargin=\oddsidemargin
\def\doublespaced{\baselineskip=\normalbaselineskip\multiply
    \baselineskip by 150\divide\baselineskip by 100}
\doublespaced

\begin{document}


\begin{flushright}
VPI-IHEP-95-08\\
TUIMP-TH-95/66\\
MSUHEP-50620\\
August, 1995
\end{flushright}
\vskip 0.4in
\centerline{\normalsize\bf
SENSITIVITY OF THE LHC TO ELECTROWEAK SYMMETRY BREAKING:}
\baselineskip=17pt
\centerline{\normalsize\bf
EQUIVALENCE THEOREM AS A CRITERION}

\vspace*{1.30cm}
\centerline{\normalsize {\bf  Hong-Jian He}~$^{(a)}$,
{}~~~~~{\bf Yu-Ping Kuang}~$^{(b)}$,~~~~~{\bf C.--P. Yuan}~$^{(c)}$ }

\vspace*{0.5cm}

\baselineskip=17pt
\centerline{\normalsize
$^{(a)}$ {\it Department of Physics and Institute of High Energy Physics}}
\baselineskip=14pt
\centerline{\normalsize\it
Virginia Polytechnic Institute and State University}
\baselineskip=14pt
\centerline{\normalsize\it Blacksburg, Virginia 24061-0435, USA}
\vspace*{0.5cm}
\centerline{\normalsize
$^{(b)}$ {\it
CCAST ( World Laboratory ), P.O.Box 8730, Beijing 100080, China }}
\baselineskip=14pt
\centerline{\normalsize\it
Institute of Modern Physics, Tsinghua University, Beijing 100084, China
\footnote{Mailing address.}}
\vspace*{0.5cm}
\centerline{\normalsize
$^{(c)}$
{\it Department of Physics and Astronomy, Michigan State University } }
\baselineskip=14pt
\centerline{\normalsize\it East Lansing, Michigan 48824, USA}

\baselineskip=18pt
\vspace{0.8cm}
\begin{abstract}

\baselineskip=17pt
\noindent
Based upon our recent study on the intrinsic connection
between the longitudinal weak-boson scatterings and
probing the electroweak symmetry breaking (EWSB) mechanism,
we reveal the profound physical content of the Equivalence Theorem (ET) as
being able to discriminate physical processes
which are sensitive/insensitive
to probing the EWSB sector.  With this physical content of the ET
as a criterion, we analyze the complete set of the bosonic
operators in the electroweak chiral Lagrangian
and systematically classify the sensitivities
to probing all these operators
at the CERN LHC via the weak-boson fusion in $W^\pm W^\pm$ channel.
This is achieved by developing
a precise power counting rule
(a generalization from Weinberg's counting method)
to {\it separately} count
the power dependences on the energy $E$ and all relevant mass scales.
\end{abstract}

\noindent
PACS number(s): 11.30.Qc, 11.15.Ex, 12.15.Ji, 14.70.--e

\normalsize
\baselineskip=19.5pt
\setcounter{footnote}{00}
\renewcommand{\thefootnote}{\alph{footnote}}
\renewcommand{\baselinestretch}{1.3}


\newpage
\noindent
{\bf 1. Introduction}
\vspace{0.3cm}

Despite the astonishing success of the Standard Model (SM) over the years,
its scalar part, the electroweak symmetry breaking (EWSB) sector,
remains as the greatest mystery.
Due to Veltman's screening theorem \cite{screening},
the current low energy data,
allowing the SM Higgs boson mass to range from $60$\,GeV to about $1$\,TeV,
tell us little about the EWSB mechanism.
It is therefore important to probe {\it all possible EWSB mechanisms,
either weakly or strongly interacting}
as long as the light Higgs particle remains undetected.

While the transverse components $V_T^a$ of $~W^{\pm},~Z^0~$
are irrelevant to the EWSB mechanism, the longitudinal weak-bosons
( $V_L^a=W^{\pm}_L,~Z^0_L~$), as the products of
the spontaneously symmetry-breaking
mechanism, are expected to be sensitive to
probing the EWSB sector. However, even for the strongly
coupled case, studying the $V_L$-scatterings does not guarantee
probing the EWSB sector in a sensitive and unambiguous way
because the spin-$0$ Goldstone bosons (GB's) are
invariant under the proper Lorentz
transformations, while, on the contrary, both $V_L$ and $V_T$
are Lorentz non-invariant
(LNI). After a Lorentz transformation, the $V_L$
 component can mix with  or even turn into a pure $V_T$.
Thus a conceptual and fundamental ambiguity arises: How can
the LNI $~V_L$-amplitudes be used to probe the
EWSB sector of which the physical mechanism should clearly be
independent of the choices of the Lorentz frames?
This motivated our recent precise
formulation of the electroweak Equivalence
Theorem (ET) in Ref.~\cite{et3}.
In the high energy region ($E\gg M_W$), the ET provides a quantitative
relation between the $V_L$-amplitude and the corresponding
GB-amplitude  \cite{et1,et2,et3};
the former is physically measurable while
the latter carries information about the EWSB sector.
Hence, the ET allows us to probe the EWSB sector
by relating it to the $~V_L$-scattering experiments.
As will be shown later, the {\it difference}
between the $V_L$- and GB-amplitudes
is intrinsically related to
the ambiguous LNI part of the $V_L$-scattering
which has the same origin as the $V_T$-amplitude,
and is thus insensitive to probing the EWSB sector.
When the LNI contributions can be safely ignored and the
Lorentz invariant (LI) scalar  GB-amplitude dominates
the experimentally measured $V_L$-amplitudes,
the physical $V_L$-scatterings can then
sensitively  and unambiguously probe the EWSB mechanism.
Since the ratio of the LNI contributions to the LI GB-amplitude
is {\it process-dependent}, it can thus determine
the sensitivities of various
scattering processes to probing the EWSB sector.

At the scale below new heavy resonances, all the effects due to
the EWSB can be parametrized by a complete set of effective operators
in the electroweak chiral Lagrangian (EWCL).
Without experimental observation of any
new light resonance in the EWSB sector, this effective
field theory approach provides an elegant way to  generally
parametrize all possible new physics effects in the low energy
region and is thus {\it complementary} to those specific model buildings.
In this paper, we take an economical and conservative viewpoint
and adopt the EWCL approach for our investigation.
We shall concentrate on
studying the bosonic operators among which
the leading order operators are universal (independent of models of
EWSB) and all the model-dependent effects are described
by the next-to-leading-order operators in the EWCL.
We show in this paper that for a given process the ratio
of the LNI contributions in the $V_L$-amplitude to the scalar GB-amplitude
varies for different effective operators.
{\it  Therefore, this ratio can be used to discriminate sensitivities
to the next-to-leading-order
effective operators as well as to the scattering processes
for probing the EWSB sector.}
The smaller this ratio, the more sensitive a process
will be to an operator.
We shall classify the sensitivities to all these effective
operators at the CERN Large Hadron Collider (LHC).
Through this analysis,
we show that the ET is not just a technical
tool in computing $V_L$-amplitudes via GB-amplitudes,
{\it as a criterion, it has an even more profound physical content for
being able to discriminate sensitivities to
different effective operators via different processes for probing the
EWSB mechanism.}

This paper is organized as follows. We first formulate the ET
as a criterion for probing the EWSB mechanism in Sec.~2, and derive
a precise electroweak power counting
rule for the EWCL formalism in Sec.~3.
Then, in Sec.~4, we classify the sensitivities of all effective
operators at the level of the $S$-matrix elements.
Finally we analyze, in Sec.~5,
the probe of the EWSB at the LHC
(a ${\rm p}{\rm p}$
collider with $\sqrt{S}=14$\,TeV) via weak-boson scatterings.
Conclusions are given in Sec.~6.
Also, a detailed analysis on the
the validity of the ET in some special kinematic regions
and its implication in probing the EWSB sector is
presented in the Appendix.

\vspace{0.5cm}
\noindent
{\bf 2. Formulating the ET as a Criterion for Probing the EWSB}
\vspace{0.3cm}

Starting from the Slavnov-Taylor identity \cite{et1,et2}
{}~~$<0\,|F^{a_1}_0(k_1)\cdots F^{a_n}_0(k_n)\,
\Phi_{\alpha}|\,0>\,=\,0 ~$\footnote{
Here, $~F_0^{a}~$ is the bare gauge fixing function and $\Phi_{\alpha}$
denotes other possible physical in/out states.}~
and making a rigorous Lehmann-Symanzik-Zimmermann (LSZ) reduction for the
external $F^a$-lines, we derived the following general identity for the
renormalized $S$-matrix elements:\footnote{
See the second paper by H.-J. He, Y.-P. Kuang and X. Li
in Ref.~\cite{et2}.}
$$
T[V^{a_1}_L,\cdots ,V^{a_n}_L;\Phi_{\alpha}]
= C\cdot T[-i\pi^{a_1},\cdots ,-i\pi^{a_n};\Phi_{\alpha}]+ B  ~~,
\eqno(2.1)                                               
$$
$$
\begin{array}{ll}
C & \equiv C^{a_1}_{mod}\cdots C^{a_n}_{mod} ~, \\
B & \equiv\sum_{l=1}^n (~C^{a_{l+1}}_{mod}\cdots C^{a_n}_{mod}
T[v^{a_1},\cdots ,v^{a_l},-i\pi^{a_{l+1}},\cdots ,
-i\pi^{a_n};\Phi_{\alpha}]\\
  & ~~~~~~~~~~~~~~+ ~{\rm permutations ~of}~
v'{\rm s ~and}~\pi '{\rm s}~)~,
  \\[0.2cm]
v^a & \equiv v^{\mu}V^a_{\mu} ~,
{}~~~~v^{\mu}~\equiv \epsilon^{\mu}_L-k^\mu /M_a = O(M_a/E) ~,
{}~~~(M_a=M_W,M_Z)~,
\end{array}
\eqno(2.1a,b,c)                                            
$$
where $~\pi^a$'s  are GB fields; and the finite
constant modification factor
$~C_{mod}^a~$ has been systematically studied in Ref.~\cite{et2},
which can be exactly
simplified as unity in some renormalization schemes \cite{hkly}.
Without losing generality \cite{et3},
let us assume that $~\Phi_\alpha~$ contains
some physical scalars, photons, or no field at all.
{}From (2.1), the LNI $~V_L$-amplitude can be decomposed
into two parts: the 1st part is
$~C\cdot T[-i\pi;\Phi_{\alpha}]~$ which is LI; the 2nd
part is the $~v_\mu$-suppressed $B$-term which is  LNI
because it contains the external {\hbox{spin-1}} $V_{\mu}$-field(s).
Such a decomposition
shows the {\it essential difference} between the $V_L$- and
the $V_T$-amplitudes. The former
contains a LI GB-amplitude that can
yield a large $V_L$-amplitude in the case of
strongly coupled EWSB sector.
We note that only the LI part of the $V_L$-amplitude
is sensitive to probing the EWSB sector,
while its LNI part contains a
significant {\it Lorentz-frame-dependent} $~B$-term
and is therefore not sensitive to the EWSB mechanism.
Thus, for a sensitive and unambiguous probe of the EWSB,
we must find conditions that the LI GB-amplitude
dominates the $V_L$-amplitude and the LNI $B$-term
can be ignored.
It is clear that one can technically improve the prediction for the
$V_L$-amplitude from the RHS of (2.1)
by including the complicated $B$-term ( or
part of $B$ ) \cite{gk},
but this is not an improvement of the equivalence between the
$V_L$- and GB-amplitudes.
 {\it The physical content of the ET is essentially
independent of how to compute the $V_L$-amplitude.
It is the LI GB-amplitude that really matters for
sensitively probing the EWSB sector.}

{}From a detailed analysis  on the LNI $V_L$-amplitude,
we can estimate the $B$-term as \cite{et3}
$$
B \approx O(\frac{M_W^2}{E_j^2})T[ -i\pi^{a_1},\cdots , -i\pi^{a_n};
  \Phi_{\alpha}] +
  O(\frac{M_W}{E_j})T[ V_{T_j} ^{a_{r_1}}, -i\pi^{a_{r_2}},
                      \cdots , -i\pi^{a_{r_n}}; \Phi_{\alpha}]~~.
\eqno(2.2)                                                   
$$
We see that the condition $~~ E_j \sim k_j  \gg  M_W , ~
(j=1,2,\cdots ,n) ~~$ {\it for each external longitudinal weak-boson} is
{\it necessary} for making the $B$-term
( and its Lorentz variation ) to be
much smaller than the GB-amplitude.
This also precisely defines the {\it safe
Lorentz frames} in which the LNI
$B$-term can be ignored, cf. (2.3). In conclusion, we give our general
and precise formulation of the ET as follows:
$$
T[V^{a_1}_L,\cdots ,V^{a_n}_L;\Phi_{\alpha}]
= C\cdot T[-i\pi^{a_1},\cdots ,-i\pi^{a_n};\Phi_{\alpha}]+
O(M_W/E_j{\rm -suppressed} ),
\eqno(2.3)                                               
$$
$$
\begin{array}{l}
E_j \sim k_j  \gg  M_W , ~~~~~(~ j=1,2,\cdots ,n ~)~~;\\
B  \ll  C\cdot T[-i\pi^{a_1},\cdots ,-i\pi^{a_n};\Phi_{\alpha}] ~~,\\
\end{array}
\eqno(2.3a,b)                               
$$
where (2.3a,b)
{\it  are the precise conditions for ignoring the LNI $B$-term
to validate the equivalence in} (2.3).
We emphasize that, in principle, the complete set of diagrams
(including those with internal
gauge boson lines) has to be considered when calculating
$~T[-i\pi^{a_1},\cdots ,-i\pi^{a_n};\Phi_{\alpha}]~$.
If not, this equivalence might not manifest for
forward or backward scatterings
for processes involving $t$- or $u$- channel diagram.
A detailed discussion on this point is given in the Appendix.

The amplitude $T$, to a finite order, can be written as
$~~T= \sum_{\ell=0}^N T_\ell
= \sum_{\ell=0}^N \bar{T}_\ell \alpha^{\ell}~$
in the perturbative calculation.
Let $~~ T_0 > T_1,\cdots, T_N \geq T_{\min}~$, where
$~T_{\min}= \{ T_0, \cdots , T_N \}_{\min}~$, then the condition (2.3b)
implies
$$
\begin{array}{ll}
B & \approx O(\frac{M_W^2}{E_j^2}) \,T_0[ -i\pi^{a_1},\cdots , -i\pi^{a_n};
   \Phi_{\alpha}] +
  O(\frac{M_W}{E_j}) \,T_0[ V_{T_j}^{a_{r_1}}, -i\pi^{a_{r_2}},
                      \cdots , -i\pi^{a_{r_n}}; \Phi_{\alpha}] \\
 & \ll  T_{\min}[-i\pi^{a_1},\cdots ,-i\pi^{a_n};\Phi_{\alpha}]  ~~.
\end{array}
\eqno(2.4)                                                 
$$
Note that {\it the above formulation of the ET discriminates processes
which are insensitive to probing the EWSB sector when either
{\rm (2.3a)} or {\rm (2.3b)} fails.}
Furthermore, {\it as a physical criterion, the condition (2.4) determines
 whether or not the corresponding $~V_L$-scattering process
in (2.3) is sensitive to probing
the EWSB sector to the desired precision in perturbative calculations.}

{}From (2.2) or the LHS of (2.4) and the precise electroweak power counting
rule (to be discussed in Sec.~3), we can easily estimate the
{\it largest} and {\it model-independent} $B$-term to be
$~~~
B_{\max} = O(g^2)f_{\pi}^{4-n}~~$ in the EWCL
formalism,\footnote{This is also true for the heavy Higgs SM.}~
which comes from the $n$-particle pure $V_L$-amplitude.
It is crucial to note that $B_{\max}$ is of the same order
of magnitude as the leading $V_T$-amplitude:
$$
B_{\max} \approx T_0[V^{a_1}_T,\cdots ,V^{a_n}_T]
= O(g^2)f_{\pi}^{4-n}~~.
\eqno(2.5)                                 
$$
Since both the largest $B$-term and the leading $V_T$-amplitude are of
$O(g^2)$~, they are therefore irrelevant to the EWSB mechanism
as pointed out in the above analysis.
Thus, {\it  (2.4) and (2.5) provide
useful criteria for discriminating physical processes
which are sensitive, marginally sensitive, or insensitive to
the EWSB sector.}

\vspace{0.5cm}
\noindent
{\bf 3. Generalized Precise Power Counting
        for the Electroweak Chiral Lagrangian}
\vspace{0.1cm}

In this section, we generalize Weinberg's counting method \cite{wei} and
develop a precise counting rule for the EWCL
in the energy region $~~M_W, m_t\ll E\ll\Lambda$,\footnote{The
generalizations of Weinberg's counting method to the
light Higgs SM and heavy Higgs SM are given in Ref.~\cite{hky}.}~
where the effective cutoff $\Lambda$  is the
upper limit of $E$ at which the EWCL formalism
ceases to be applicable.
In this work we shall assume that
the EWSB sector does not contain any new resonance below the scale
$~~\Lambda\simeq 4\pi f_\pi \simeq 3.1$\,TeV.~
We want to {\it separately} count the power dependences of the amplitudes
on the energy $E$, the cutoff scale
$\Lambda$ of the EWCL and the Fermi
scale $f_\pi = 246$\,GeV ($\sim M_W, m_t$).\footnote{This is
essentially different from the previous counting for the
heavy Higgs SM where only the {\it sum} of the powers of
$E$ and $m_H$ has been counted \cite{hveltman}.}~
This is {\it crucial} for estimating the order of magnitude of an amplitude
at any given order of perturbative calculation.
For example, an amplitude of $~O(E^2/f_\pi^2)~$ differs by an order
of magnitude
from an amplitude of $~O(E^2/\Lambda^2)~$ in spite that they have the same
$E$-dependence.
Since the weak-boson mass $M_W=g f_\pi/2$ and the fermion mass
$m_f=y_f f_\pi/\sqrt{2}$, we can
count them in powers of the coupling constants $g$ and $y_f$
and the vacuum expectation value $f_\pi$.
The $SU(2)$ weak gauge
coupling $g$ and the top quark Yukawa coupling $y_t$ are close to $1$
and thus will not significantly affect the order of magnitude estimates.
The electromagnetic $U(1)_{em}$ coupling
$~e=g\sin\theta_W~$ is smaller than $g$ by
a factor of 2. The Yukawa couplings of all
light SM fermions other than the top quark are negligibly small.
In our following precise counting rule, the dependences on coupling
constants $~g,~g^{\prime} ({\rm or}~ e)~$ and $~y_t~$ are included,
while all the light
fermion Yukawa couplings [ $y_f~(\neq y_t) \ll 1$ ] are ignored.

The original Weinberg's power counting rule
was derived only for counting the energy dependence
in the un-gauged nonlinear $\sigma$-model as a description of low energy
QCD interaction~\cite{wei}.
The {\it general} features of Weinberg's counting method are:
(i). The total dimension $~D_T~$ of
an $S$-matrix element $~T~$ is determined
by the number of external lines and the space-time dimension;
(ii).   Assume that all mass poles
in the internal propagators of $~T~$ are much smaller
than the typical energy scale $E$ of $~T~$, then the total
dimension $~D_m~$ of the $E$-independent coupling constants included in
$~T~$ can be directly counted
according to all types of vertices it contains.
Hence, the total $E$-power $~D_E~$ for
$~T~$ is given by $~~D_E= D_T -D_m~~$.

Here, we shall make a natural generalization of
Weinberg's power counting method
for the EWCL in which, except the light SM
gauge bosons, fermions and would-be GB's, all possible heavy fields have
been integrated out.
It is clear that in this case the above
conditions (i) and (ii) are satisfied.
The total dimension of an $L$-loop $S$-matrix element $T$ is
$$
D_T = 4-e~~,
\eqno(3.1)                                          
$$
where $~~e=e_B+e_F~~$, and $e_B$ ($e_F$) is
 the number of external bosonic
(fermionic) lines. Here the dimensions of the external spinor wave
functions are already included in $D_T$.
For external fermionic lines, we only
count the SM fermions with masses
$~~m_f\leq m_t \sim O(M_W)\ll E ~$. So the
spinor wave function of each external fermion
will contribute an energy factor
$~E^{1/2}~$ for $~~E\gg m_f~$, where the spinor wave functions are
normalized as
$~~ \bar{u}(p,s)u(p,s^\prime )=2m_f\delta_{ss^\prime}~~$, etc.
Let us label the different types of vertices by an index $n$. If
the vertex of type $n$ contains $b_n$ bosonic lines,
$f_n$ fermionic lines and $d_n$ derivatives,
then the dimension of the $~E$-independent
effective coupling constant in $T$ is
$$
D_m= \displaystyle\sum_n {\cal V}_n (4-d_n-b_n-\frac{3}{2}f_n) ~~,
\eqno(3.2)                              
$$
where ${\cal V}_n$ is the number of vertices of type $n$.
Let $i_B$ and $i_F$ be the numbers of
internal bosonic and fermionic lines,
respectively. ( $i_B$ also includes possible internal ghost lines.)
Define  $~~i=i_B+i_F~~$,
we have, in addition, the following general relations
$$
\begin{array}{l}
\displaystyle\sum_n b_n{\cal V}_n =2i_B+e_B~~,~~~~
\displaystyle\sum_n f_n{\cal V}_n =2i_F+e_F ~~,~~~~
L=1+i-\displaystyle\sum_n {\cal V}_n ~~.
\end{array}
\eqno(3.3)
$$                                          
Among the external vector-boson lines, each $V_L$-line contains a
polarization vector $\epsilon_L$
which is of $O(E/M_W)$, and each $v_\mu$
defined in (2.1c) is of $O(M_W/E)$. Let $e_L$ and  $e_v$ denote the
numbers of external $V_L$ and $v_\mu$ lines,
respectively. Then from (3.1),
(3.2) and (3.3), the leading energy power in $T$ is
$$
D_E  ~= D_T-D_m+e_L-e_v
     ~= 2L +2 + \sum_n {\cal V}_n(d_n +\frac{1}{2}f_n-2)+ e_L- e_v ~~.
\eqno(3.4)                                                     
$$
This is just the Weinberg's counting rule \cite{wei}
in its generalized form
with the gauge boson,
ghost and fermion fields and possible $v_\mu$-factors
included.\footnote{(3.4) is clearly valid for
any gauge theory satisfying the above conditions (i) and (ii).}

A subtle point should be noted. To show this, we
take the $V_LV_L \rightarrow V_LV_L$
scattering amplitude as an example, in which $e_L=4$
and $e_v=e_F=0$.
To the lowest order of the EWCL, the leading
powers of $E$ in $T[V^{a_1}_L, \cdots ,V^{a_4}_L]$ and
$T[\pi^{a_1}, \cdots ,\pi^{a_4}]$ are $E^4$
and $E^2$, respectively. This tells us that the naive power counting for
$V_L$-amplitude only gives the leading $E$-power for individual graphs.
 It does not reflect the fact that gauge invariance
causes the cancellations of the $E^4$-terms, and leads to the
final $E^2$-dependence of the whole $V_L$-amplitude.
Thus the naive power counting
of the $V_L$-amplitudes does not give the correct answer. However,
 the power counting of GB-amplitude does
give the correct $E$-dependence because, unlike in the $V_L$-amplitudes,
 there is
generally no large $E$-power cancellations in the GB-amplitudes.
Therefore based on our ET identity (2.1)
the correct counting in powers of $E$ for the $V_L$-amplitude
can be given by counting the corresponding GB-amplitude plus
the $B$-term.
So in what follows, we shall not directly count the $E$-dependence
in diagrams with
external longitudinal weak-boson lines. They will be counted
through counting the RHS of the ET identity in (2.1).
We shall therefore drop the $e_L$ term in (3.4),
and {\it make the convention that the number of external
vector-boson lines $e_V$ counts only the
number of external $V_T$-lines and photon lines.}

In the following, we further
develop a precise power counting rule for the EWCL
to separately count the dependence of $S$-matrix elements on energy ($E$),
 cutoff scale ($\Lambda$) of the effective Lagrangian
and vacuum expectation value ($f_\pi$).
This separate counting on the powers of
$E$, $\Lambda$ and $f_\pi$ is important for
estimating contributions to scattering amplitudes
from various effective operators in the Lagrangian.
In general, the EWCL can be constructed as \cite{app,peccei}:
$$
\begin{array}{ll}
{\cal L}_{eff}&
= \displaystyle\sum_n
\ell_n\displaystyle\frac{f_\pi~^{r_n}}{\Lambda^{a_n}}
{\cal O}_n(W_{\mu\nu},B_{\mu\nu},D_\mu U,U,f,\bar{f})
= {\cal L}_G + {\cal L}_{S} + {\cal L}_F
\end{array}
\eqno(3.5)                          
$$
where
$$
\begin{array}{l}
D_{\mu}U =\partial_{\mu}U + ig{\bf W}_{\mu}U -ig^{\prime}U{\bf B}_{\mu}~,
\\[0.25cm]
U=\exp [i\tau^a\pi^a/f_\pi ]~,~~~
{\bf W}_{\mu}\equiv W^a_{\mu}\displaystyle\frac{\tau^a}{2}~,~~~
{\bf B}_{\mu}\equiv B_{\mu}\displaystyle\frac{\tau^3}{2}~.\\[0.20cm]
\end{array}
$$
$f$($\bar{f}$) is the SM fermion with mass $~m_{f}\leq
O(m_t)\simeq O(M_W)~$.
$~{\cal L}_G~$, $~{\cal L}_{S}~$ and $~{\cal L}_F~$ denote
 gauge boson kinetic terms, scalar boson interaction terms
(containing GB self-interactions and gauge-boson-GB interactions),
and fermion interaction terms, respectively.
For clearness, we have factorized out the dimensionful parameters
$~f_\pi~$ and $~\Lambda~$ in the coefficients so that the dimensionless
factor $~~\ell_i\sim O(1)$.\footnote{This makes our definitions of the
$~\ell_i$'s different from the $~\alpha_i$'s in Ref.~\cite{app}
by a factor of $~(f_{\pi}/\Lambda )^2~$.}~
We note that  $~f_\pi~$  and $~\Lambda~$
are the two essential scales in any effective Lagrangian
that describes the spontaneously broken symmetry.  The former determines the
symmetry breaking scale while the latter determines the scale at which
new resonance(s) besides the light fields
(such as the SM
weak bosons, would-be Goldstone bosons and fermions) may appear.
For the non-decoupling scenario,
the effective cutoff scale $\Lambda$ cannot be arbitrarily large.
It is $~~\Lambda = \min ( M_{SB}, 4\pi f_\pi )~~$,
where $~M_{SB}~$ is the mass of the lightest new resonance,
and $~~\Lambda \simeq 4\pi f_\pi~$ \cite{georgi} for the case without
new resonance in the EWSB sector.
In (3.5), $~~ r_n=4+a_n-D_{{\cal O}_n} ~$, where $~D_{{\cal O}_n}
={\rm dim}({\cal O}_n)~$.
For the bosonic part of EWCL, we have \cite{app}:

$$
\begin{array}{ll}
{\cal L}_G &=~ -\frac{1}{2}{\rm Tr}({\bf W}_{\mu\nu}{\bf W}^{\mu\nu})
              -\frac{1}{4}B_{\mu\nu}B^{\mu\nu}  ~~,\\[0.25cm]
{\cal L}_{S} &
 = {\cal L}^{(2)}+{\cal L}^{(2)\prime}+
             \displaystyle\sum_{n=1}^{14} {\cal L}_n ~~,\\[0.25cm]
{\cal L}^{(2)} & =
   \frac{f_\pi^2}{4}{\rm Tr}[(D_{\mu}U)^\dagger(D^{\mu}U)]   ~~,\\[0.20cm]
{\cal L}^{(2)\prime} & =\ell_0 (\frac{f_\pi}{\Lambda})^2~\frac{f_\pi^2}{4}
               [ {\rm Tr}({\cal T}{\cal V}_{\mu})]^2 ~~,\\[0.2cm]
{\cal L}_1 & = \ell_1 (\frac{f_\pi}{\Lambda})^2~ \frac{gg^\prime}{2}
B_{\mu\nu} {\rm Tr}({\cal T}{\bf W^{\mu\nu}}) ~~,\\[0.2cm]
{\cal L}_2 & = \ell_2 (\frac{f_\pi}{\Lambda})^2 ~\frac{ig^{\prime}}{2}
B_{\mu\nu} {\rm Tr}({\cal T}[{\cal V}^\mu,{\cal V}^\nu ]) ~~,\\[0.2cm]
{\cal L}_3 & = \ell_3 (\frac{f_\pi}{\Lambda})^2 ~ig
{\rm Tr}({\bf W}_{\mu\nu}[{\cal V}^\mu,{\cal V}^{\nu} ]) ~~,\\[0.2cm]
{\cal L}_4 & = \ell_4 (\frac{f_\pi}{\Lambda})^2
                     [{\rm Tr}({\cal V}_{\mu}{\cal V}_\nu )]^2 ~~,\\ [0.2cm]
{\cal L}_5 & = \ell_5 (\frac{f_\pi}{\Lambda})^2
                     [{\rm Tr}({\cal V}_{\mu}{\cal V}^\mu )]^2 ~~,\\  [0.2cm]
{\cal L}_6 & = \ell_6 (\frac{f_\pi}{\Lambda})^2
[{\rm Tr}({\cal V}_{\mu}{\cal V}_\nu )]
{\rm Tr}({\cal T}{\cal V}^\mu){\rm Tr}({\cal T}{\cal V}^\nu) ~~,\\[0.2cm]
{\cal L}_7 & = \ell_7 (\frac{f_\pi}{\Lambda})^2
[{\rm Tr}({\cal V}_\mu{\cal V}^\mu )]
{\rm Tr}({\cal T}{\cal V}_\nu){\rm Tr}({\cal T}{\cal V}^\nu) ~~,\\[0.2cm]
{\cal L}_8 & = \ell_8 (\frac{f_\pi}{\Lambda})^2~\frac{g^2}{4}
[{\rm Tr}({\cal T}{\bf W}_{\mu\nu} )]^2  ~~,\\[0.2cm]
{\cal L}_9 & = \ell_9 (\frac{f_\pi}{\Lambda})^2 ~\frac{ig}{2}
{\rm Tr}({\cal T}{\bf W}_{\mu\nu}){\rm Tr}
        ({\cal T}[{\cal V}^\mu,{\cal V}^\nu ]) ~~,\\[0.2cm]
{\cal L}_{10} & = \ell_{10} (\frac{f_\pi}{\Lambda})^2\frac{1}{2}
[{\rm Tr}({\cal T}{\cal V}^\mu){\rm Tr}({\cal T}{\cal V}^{\nu})]^2 ~~,\\[0.2cm]
{\cal L}_{11} & = \ell_{11} (\frac{f_\pi}{\Lambda})^2
{}~g\epsilon^{\mu\nu\rho\lambda}
{\rm Tr}({\cal T}{\cal V}_{\mu}){\rm Tr}
({\cal V}_\nu {\bf W}_{\rho\lambda}) ~~,\\[0.2cm]
{\cal L}_{12} & = \ell_{12}(\frac{f_\pi}{\Lambda})^2 ~2g
                    {\rm Tr}({\cal T}{\cal V}_{\mu}){\rm Tr}
                  ({\cal V}_\nu {\bf W}^{\mu\nu}) ~~,\\[0.2cm]
{\cal L}_{13} & = \ell_{13}(\frac{f_\pi}{\Lambda})^2~
      \frac{gg^\prime}{4}\epsilon^{\mu\nu\rho\lambda}
      B_{\mu\nu} {\rm Tr}({\cal T}{\bf W}_{\rho\lambda}) ~~,\\[0.2cm]
{\cal L}_{14} & = \ell_{14} (\frac{f_\pi}{\Lambda})^2~\frac{g^2}{8}
\epsilon^{\mu\nu\rho\lambda}{\rm Tr}({\cal T}{\bf W}_{\mu\nu})
{\rm Tr}({\cal T}{\bf W}_{\rho\lambda})   ~~,\\[0.2cm]
\end{array}
\eqno(3.6)                                         
$$
where
$~{\bf W}_{\mu\nu} =\partial_{\mu}{\bf W}_{\nu}-\partial_{\nu}{\bf W}_{\mu}
                      -ig [{\bf W}_\mu , {\bf W}_\nu ] ~$,
$~{\cal V}_{\mu}\equiv (D_{\mu}U)U^\dagger~$,
and $~{\cal T}\equiv U\tau_3 U^{\dag} ~$.
There are fifteen next-to-leading-order effective operators among which
there are twelve $CP$-conserving operators
($~{\cal L}^{(2)\prime},{\cal L}_{1\sim 11}~$) and three $CP$-violating
operators (~${\cal L}_{12\sim 14}~$).
Furthermore, the operators $~{\cal L}_{6,7,10}~$
violate custodial $SU(2)_C$ symmetry
(~even after $g^{\prime}$ being turned off~) contrary to
 $~{\cal L}_{4,5}~$ which contain $SU(2)_C$-invariant
 pure GB interactions.
The coefficients ( $\ell_n$'s ) of all the above operators are
model-dependent and carry
information about possible new physics beyond the SM.
The dimension-$2$ custodial $SU(2)_C$-violating operator
$~{\cal L}^{(2)\prime} ~$  has a coefficient of
$~O((f_\pi/\Lambda )^2 )~$ since it is  proportional to
$~\delta\rho \simeq O(m_t^2/(16\pi^2 f_\pi^2))
\simeq O((f_\pi /\Lambda )^2 )~$ for the top Yukawa coupling being
of $~O(1)$.

In the non-decoupling scenario \cite{georgi,review},
all the coefficients for dimension-$4$ operators are suppressed
by a factor $~(f_\pi/\Lambda )^2\simeq 1/(16\pi^2)~$
because they arise from
 the derivative expansion in $~(D_\mu/\Lambda )^2~$.
After the small $CP$-violating effects from the
Cabibbo-Kobayashi-Maskawa mixings are ignored in the lowest order
fermionic operators contained in $~{\cal L}_F~$,
all the one-loop level new divergences generated from
$~{\cal L}_G+{\cal L}_F+{\cal L}^{(2)}~$ are thus $CP$-invariant.
Therefore, the $CP$-violating operators $~{\cal L}_{12\sim 14}~$ are
actually {\it decoupled} at this level, and their
coefficients can have values significantly larger or smaller than that from
the naive dimensional analysis \cite{georgi}.
Since the true mechanism for $CP$-violation remains un-revealed,  we
shall consider in this paper the coefficients $~\ell_{12\sim 14}~$
to be around of $~O(1)~$.

Consider the $S$-matrix element $T$ at $L$-loop order.
Since we are dealing
with a spontaneously broken gauge theory which possesses
 a nonvanishing vacuum expectation value ( $f_\pi$ ),
 $T$ can always be written as $~f_{\pi}^{D_T}~$
times some dimensionless function
of $~E,~\Lambda$, and $f_\pi~$, etc.
The $E$-power dependence has been given by our generalized Weinberg
formula (3.4).
Since the cutoff scale $\Lambda$ in the EWCL is much
larger than $f_\pi$ (for ~$\Lambda/f_\pi\simeq 4 \pi$),
it is crucial to separately count $\Lambda$ and $f_\pi$
to correctly estimate the magnitude of an amplitude.
The $~\Lambda$-dependence in $T$ can only come from two sources:
\begin{description}
\item{(i).} From tree vertices:
$T$ contains $~{\cal V}=\displaystyle\sum_n {\cal V}_n~$ vertices,
each of which contributes a factor
$~1/\Lambda^{a_n}~$ so that the total factor from ${\cal V}$-vertices is
$~1/\Lambda^{\sum_n a_n}~$;
\item{(ii).}  From loop-level: Since each loop brings
a factor $~(1/4\pi)^2 \simeq (f_{\pi}/\Lambda )^2~$, the total
$~\Lambda$-dependence from loop contribution is $~~1/\Lambda^{2L}~~$.
\end{description}
Hence the total $~\Lambda$-dependence given by the above two sources
is $~~ 1/\Lambda^{\sum_na_n +2L}~~$.
{}From the above discussion, we conclude the following
precise counting rule for $~T~$:
$$
\begin{array}{l}
T= c_T f_\pi^{D_T}\displaystyle
\left(\frac{f_\pi}{\Lambda}\right)^{N_{\cal O}}
\left(\frac{E}{f_\pi}\right)^{D_{E0}}\left(\frac{E}{\Lambda}\right)^{D_{EL}}
\left(\frac{M_W}{E}\right)^{e_v} H(\ln E/\mu)~~,\\[0.5cm]
N_{\cal O}=\displaystyle\sum_n a_n~,~~~~
D_{E0}=2+\displaystyle\sum_n {\cal V}_n(d_n+\frac{1}{2}f_n-2)~, ~~~~
D_{EL}=2L~,\\
\end{array}
\eqno(3.7)                                                  
$$
where the dimensionless coefficient $~c_T~$ contains possible powers of
gauge couplings ($~g,g^\prime~$)
and Yukawa couplings ($~y_f~$) from the vertices
in $~T~$.
$~H$ is a function of $~\ln (E/\mu )~$ which is insensitive to $E$.
(Here $\mu$ denotes the relevant renormalization scale
for loop calculations.)
We note that because  pure GB vertices contain
the highest power of derivatives
at each order of derivative expansion, (3.7) shows that
the leading $E$-power dependence is
always given by pure GB self-interacting
graphs. The same conclusion holds for pure $V_L$-scattering amplitudes
since they can be decomposed into the corresponding
GB-amplitudes plus the $~M_W/E$-suppressed $B$-term  [cf. (2.1)].

\vspace{0.5cm}
\noindent
{\bf 4. Classification of sensitivities at the
level of $S$-matrix elements}
\vspace{0.3cm}

Armed with the above counting rule (3.7),
we can easily estimate the contributions
from various effective operators in the EWCL to
any scattering process such that we can systematically classify
the sensitivities to the next-to-leading-order effective operators
for probing the EWSB sector at the LHC.
Our electroweak power counting analysis
makes it possible to quickly grasp the overall
semi-quantitative physical picture
which provides useful guidance
on selecting relevant operators and scattering processes to perform
further detailed studies.

In this paper we shall concentrate on the high
energy weak-boson scatterings.
As shown in Ref.~\cite{wwlhc}, for the non-resonance case
(i.e. no new light resonance in the symmetry breaking sector)
the most important scattering process
for probing the EWSB sector is the same-charged channel:
$~W^{\pm}W^{\pm}\rightarrow W^{\pm}W^{\pm}~$.
In Tables~1a and 1b we
estimate the contributions from the lowest order
(model-independent) operators in
$~{\cal L}_0\equiv {\cal L}_G +{\cal L}_F +{\cal L}^{(2)}~$ up to one-loop
and from all the next-to-leading-order
(model-dependent) bosonic operators in (3.6) at tree-level
for $~W^{\pm}W^{\pm}\rightarrow W^{\pm}W^{\pm}~$.
For instance, the commonly discussed operators $~{\cal L}_{4,5}~$
contribute the model-dependent leading term of
$~~O(\frac{E^2}{f_\pi^2}\frac{E^2}{\Lambda^2})~~$
to the $~T[4W_L]~$ amplitude,
and the sub-leading term of
$~~O(g\frac{E}{f_\pi}\frac{E^2}{\Lambda^2})~~$ to
the $~T[3W_L,W_T]~$ amplitude.
The model-independent
and model-dependent contributions to various $B$-terms
are summarized in Tables~2a and 2b, in which
$B^{(i)}_{\ell}$ ($i=0,\cdots ,3;~\ell =0,1,\cdots$)
denotes the $B$-term from $V_L$-amplitudes containing $0,1,2,3$ external
$V_T$-lines, respectively.
Here $B^{(i)}_0$ is obtained from the tree level and
$B^{(i)}_1$ from the next-to-leading order.
We see that the largest $B$-term is
$B^{(0)}_0$ from the $4W_L$ amplitudes, as given in (2.5).
The $B^{(0)}_0$, which is $O(g^2)$, is
a model-independent constant containing only
the SM gauge coupling constants.
All the other $B$-terms are further suppressed
by a factor of $M_W/E$ or $(E/\Lambda )^2$, or a product of them.

{}From Tables~1$\sim$2 and our recent exhaustive study \cite{hky}
for  $~V^aV^b\rightarrow V^cV^d~$  scatterings,
we  further classify in Table 3
the sensitivities to all the bosonic  operators for probing
the EWSB sector
directly (from pure GB interactions) or indirectly (from interactions
suppressed by the SM gauge coupling constants).
The classification is based on the following hierarchy in the power
counting:
$$
\frac{E^2}{f_\pi^2}\gg \frac{E^2}{f_\pi^2}\frac{E^2}{\Lambda^2},
{}~g\frac{E}{f_\pi} \gg g\frac{E}{f_\pi}\frac{E^2}{\Lambda^2}, ~g^2
\gg g^2\frac{E^2}{\Lambda^2}, ~g^3\frac{f_\pi}{E}
\gg g^3\frac{Ef_\pi}{\Lambda^2} \gg g^4\frac{f_\pi^2}{\Lambda^2}~~.
\eqno(4.1)                               
$$
In the TeV region, for
$~~E\in (750\,{\rm GeV},~1.5\,{\rm TeV})$,
this gives:
$$
\begin{array}{l}
(9.3,37)\gg (0.55,8.8),(2.0,4.0)\gg (0.12,0.93),(0.42,0.42) \gg \\
 ~~~~(0.025,0.099),(0.089,0.045)\gg (5.3,10.5)\times 10^{-3}
\gg (1.1,1.1)\times 10^{-3} ~~,
\end{array}
\eqno(4.2)                             
$$
where $E$ is taken to be the invariant mass of the $VV$ pair.
The numerical values in (4.2) convincingly show the existence of
the power counting hierarchy in (4.1).
This determines the order of magnitude
of all precise results from detailed calculations.
This  hierarchy makes it possible  to
classify the sensitivities of various scattering processes to
the complete set of the effective operators in the EWCL.
The construction of this power counting hierarchy
can be understood as follows.
The leading term $~\frac{E^2}{f_\pi^2}~$ in (4.1) comes from the
model-independent lowest order $4V_L$ ($\neq 4Z_L$) scatterings.
 Starting from this leading term,
(4.1) is built up by {\it increasing either
the number of derivatives
(i.e. the power of $E/\Lambda$) or the number of
external transverse gauge bosons
(i.e. the power of gauge coupling constants).}
The next-to-leading-order contributions from the derivative expansion are
always suppressed
by $~E^2/\Lambda^2~$ relative to the model-independent leading term.
Also, when each external $V_L$-line is replaced
by the corresponding $V_T$-line,
a factor $\frac{E}{f_\pi}$ in the amplitude would be replaced by
a gauge coupling $g$ (or $g^\prime$).\footnote{
The counting on the amplitude $T[4W_T]$ is an exception of this rule
because it can have a contribution from the vector-boson kinetic term.
This exception can be found at the upper-right-hand corner of
Table~1a.}~  This explains why the power counting hierarchy takes
the form of (4.1).

Table~3 is organized in accord with the power counting hierarchy
given in (4.1) for $VV$ scattering amplitudes.
It shows the {\it relevant} effective new physics operators and the
corresponding physical processes for probing the EWSB sector
when calculating the scattering amplitudes to the required precision.
For instance, the model-independent operator $~{\cal L}_0~$
can be probed
via studying the leading tree-level scattering amplitude
$~T[4V_L]~(\neq T_0[4Z_L])~$ which is of $~O(\frac{E^2}{f_\pi^2})~$.
To test the model-dependent operators $~{\cal L}_{4,5,6,7,10}~$ demands
a higher precision than the leading tree level contribution by a factor
of $~\frac{E^2}{\Lambda^2}~$.
For examples, in the high energy region, the $~4V_L$ scatterings can
sensitively probe $~{\cal L}_{4,5}~$, while $~{\cal L}_{6,7}~$ can
be probed via $~2W_L+2Z_L~$ or $~4Z_L~$ scattering and $~{\cal L}_{10}~$
can only be tested  via $~4Z_L~$ scattering.
But, as shown in Table~3, to probe the operators
$~{\cal L}_{2,3,9,10,11;12}~$, one has to detect
the $~3V_L+V_T$
scatterings, which are further suppressed by a factor $~\frac{M_W}{E}~$
relative to the leading model-dependent contributions from the
$~{\cal L}_{4,5}~$ and $~{\cal L}_{6,7,10}~$ via $4V_L$ processes.
Since the model-independent
leading order $2V_T+2V_L$ and $4V_T$ amplitudes
(from ${\cal L}_0$) and the largest constant
$B$-term ($B^{(0)}_0$)  are all of the same order,
i.e. $~O(g\frac{E}{f_\pi}\frac{E^2}{\Lambda^2},g^2)~$
[cf. (4.2)],\footnote{
They can in principle be separated if the polarization of
the external $V$-lines are identified.
For the final state $V$'s, one can study the angular
distribution of the leptons from $V$-decay.
For the incoming $V$'s, one can use forward-jet
tagging and central-jet vetoing to select
longitudinal $V$'s \cite{wwww}.}~
it requires  a significantly higher precision
to sensitively probe these operators which can only contribute
the $g$-suppressed indirect EWSB information
and therefore are more difficult to be tested.
Finally, the operators $~{\cal L}_{1,8;13,14}~$ can be probed
via the amplitude
$~T_1[2V_L,2V_T] ~(\neq T_1[2Z_L,2Z_T])~$ which is of
$~O(g^2\frac{E^2}{\Lambda^2},g^3\frac{f_\pi}{E})~$ and numerically much
smaller [cf. (4.2)].
Therefore, $~{\cal L}_{1,8;13,14}~$
should be effectively probed
via scattering processes other than the $VV$-fusion,
for instance, via $~q\bar{q}\rightarrow VV$.

In summary, applying the power counting technique allows us to
conveniently estimate contributions of various operators
to any scattering amplitude.
For a given scattering process, this result tells us which
operators can be sensitively probed. Similarly, the same result can
also tells us which process would be most sensitive for probing
new physics via a given effective operator.
In the next section, we shall examine the $W^+W^+ \rightarrow W^+W^+$
process at the LHC to illustrate how to use
the electroweak power counting method
to estimate the event rates and
how to use the ET as a physical criterion
to classify the sensitivity of this scattering process
to the next-to-leading order
bosonic operators in the EWCL.

\vspace{0.5cm}
\noindent
{\bf 5. Probing EWSB Mechanism at the LHC via Weak-Boson Scatterings}
\vspace{0.3cm}

In this section, we shall study the production rate of
$W^+W^+ \rightarrow W^+W^+$ at the LHC.
To calculate the event rate, we multiply
the luminosity of the incoming weak-boson pair $VV$
(obtained by using the
effective-$W$ approximation \cite{effective-W})
and the constituent cross section of the weak-boson scattering
(derived from the amplitude which has been
estimated by our power counting analysis in the last section).
Note that the validity of the effective-$W$ approximation
requires the $VV$ invariant mass $~~M_{VV}\gg 2M_W~~$ \cite{effective-W},
which coincides with the condition in (2.3a) for ignoring
the LNI $B$-term to apply the ET.\footnote{ Here, we have
reasonably taken the typical energy scale $E$ of the $VV$ scattering
to be $M_{VV}$ to estimate the event rates.}~
Thus, the effective-$W$ approximation and the ET
have similar precisions in computing the
event rate from $V_LV_L$ fusion process in hadron collisions.
As $M_{VV}$ increases, they become more accurate.
It is known that the effective-$W$ approximation is less accurate for
sub-processes involving transverse gauge bosons.
Generally speaking, a factor of $2$ to $5$ uncertainty in its rate
is understood \cite{gunion}.
Nevertheless, the effective-$W$ method has
been widely used in the literature for calculating event rates from
gauge-boson (either transversely or longitudinally
polarized) fusion processes because it is easy to implement and can be
used to reasonably estimate event rates before any exact calculation is
available.
As to be shown shortly, our power counting analysis
for the constituent cross section agrees well with explicit calculation
within a factor of $~2$. Hence,
it is appropriate to apply the power counting analysis
together with the effective-$W$ approximation
for estimating the event rates from weak-boson fusion at
the LHC. When applying our power counting analysis,
we have reasonably ignored the angular dependence in the scattering
amplitudes (cf. Tables~1$\sim$2)
because it will not affect the order of magnitude estimates
for the total cross sections (or the event rates).

Let us denote the production rate for the scattering
process $~W_{\alpha}^+W_{\beta}^+\rightarrow W_{\gamma}^+W_{\delta}^+~$
as $~R_{\alpha\beta\gamma\delta (\ell)}~$,
where $\alpha ,\beta ,\gamma ,\delta =L,T$ label the polarizations
of the $W$-bosons
and $~\ell=0,1,\cdots~$ indicates contributions from
tree, $1$-loop, $\cdots$, respectively.
Up to the one-loop level, we define
$$
\begin{array}{l}
R_{\alpha\beta\gamma\delta}
=R_{\alpha\beta\gamma\delta (0)} +R_{\alpha\beta\gamma\delta (1)} ~~,
\\[0.2cm]
 R_{\alpha\beta\gamma\delta (\pm )}
=R_{\alpha\beta\gamma\delta (0)} \pm
|R_{\alpha\beta\gamma\delta (1)}| ~~.
\end{array}
\eqno(5.1a,b)
$$
Also, we use $~R_B~$ to denote the rate contributed by the
largest $B$-term in the $VV \rightarrow VV$ scatterings, which
is $O(g^2)$, cf. (2.5).
For convenience, we use the subscript ``$_S$'' to stand for
summing up the polarizations of the corresponding gauge boson.

To check the reliability of our power counting method,
we have compared our results for the $W^+W^+$ scatterings
with those in Fig.~8 of Ref.~\cite{BDV}
in which all the initial state polarizations
of the weak-bosons were summed over.\footnote{We
have adopted the same effective-$W$
approximation as Ref.~\cite{BDV}.}~
As shown in Fig.~1, both results coincide
well within a factor of $2$.
This is a convincing example showing that
the semi-quantitative physical picture
can be quickly grasped by our power counting analysis without performing
complicated precise numerical calculations.

In Fig.~2a we give our power counting estimates
for the LHC production rates
of the $W^+_L W^+_L$ pairs
from different polarizations of the initial state $W$-bosons.
In this plot, we did not include
any finite part of contributions from
the next-to-leading-order operators by setting
the renormalized coefficients $~\ell_{0\sim 14}$ to be
zero.\footnote{It is understood that the divergent pieces
from one-loop calculations have been absorbed by the
coefficients of the corresponding
next-to-leading-order effective operators
\cite{app,georgi}.}
As clearly shown in Fig.~2a, the rate from $4W_L$ scattering
dominates and the rate from $W_T+3W_L$ scattering is lower by about
an order of magnitude for large $M_{WW}$ in spite of the fact that
the $W_T W_L$ luminosity is larger than the $W_LW_L$
luminosity in the initial state.
Also separately shown in the same figure is the
event rate $~|R_B|~$ contributed by the
largest $B$-term [cf. (2.1) and (2.5)]
which is even significantly lower than that from the $W_T+3W_L$ scattering
by a factor of $2\sim 7$ for $~M_{WW}>500$\,GeV.
However, the rate from $W_TW_T$ initial state is lower than
that from the $B$-term in the $4W_L$ amplitude as $~E\geq 600$\,GeV.
This implies that if the contribution from $W_TW_T$ initial state
is to be included in calculating the total production rate
of the  $W_LW_L$ pair,
the contribution from the $B$-term in the $4W_L$ amplitude
also has to be included because
they are of the same order in magnitude.
If, however, only the pure Goldstone boson amplitude
$~T[\pi^+ \pi^+ \rightarrow \pi^+ \pi^+]~$
is used to calculate the $4W_L$-amplitude
(with the $B$-term ignored)
the contribution from $~T[W^+_T W^+_T \rightarrow W_L^+ W_L^+]~$
should also be consistently ignored for computing the
total rate of $W^+_L W^+_L$ pair production
via the weak-boson fusion mechanism.

As shown in Ref.~\cite{wwlhc}, it is possible to statistically,
though not on the event-by-event basis,
choose event with longitudinally polarized
$W$-bosons in the initial state
by applying the techniques of
forward-jet tagging and central-jet vetoing.
In this work we do not intend to study the details of the
event kinematics, and
we shall sum over all the initial state polarizations
for the rest of discussions.
Let us first compare the rates for different polarizations in the final
state. Fig.~2b
shows that the rate of $W_LW_L$ final state dominates, while the
rate of $B$-term and the
rates of $W_LW_T$ and $W_TW_T$ final states are of the same order,
and all of them are about an $O(10)$ to $O(10^2)$
lower than the rate of $W_LW_L$ final state in the energy region
$~E=M_{WW} > 500\,{\rm GeV}$.
This makes it clear that if one wants to increase the precision
in calculating the total event rates by
including the small contribution from
the $B$-term in pure $4W^+_L$ scattering,
then the contributions from $W^+_SW^+_S\rightarrow W^+_TW^+_T$
and $W^+_SW^+_S\rightarrow W^+_LW^+_T$ scatterings should also be
consistently included. Otherwise, they must all
be neglected together.
Hence, from Figs.~2a and 2b, we conclude that the scattering process
$W^+_L W^+_L \rightarrow W^+_L W^+_L$ dominates the $W^+W^+$-pair
productions when all
the model-dependent coefficients $~\ell_{0\sim 14}$
in (3.6) are set to be zero.

For nonvanishing  $~\ell_{0\sim 14}$,
we classify the sensitivities to all the next-to-leading-order
bosonic operators at the LHC
via the scattering process $W^+W^+ \rightarrow W^+W^+$.
Our criterion for discriminating different sensitivity levels
(sensitive, marginally sensitive, or insensitive)
to probe a particular operator via the production of $W^+W^+$ pairs
is to compare its contribution to the event rate
( $|R_{\alpha\beta\gamma\delta (1)}|$ ) with that from the
largest model-independent contribution of the LNI $B$-term ( $|R_B|$ ).
Without knowing the values of the {\it model-dependent coefficients}
($\ell_i$'s), we show in Figs.~3$\sim$4 the results
for varying $~|\ell_i|~$ from $~O(1)$ to $~O(10)$.
Here, the polarizations of the initial and the final states
have been summed over.
In Figs.~3a and 3b, we consider the coefficients ( $\ell_i$'s )
to be naturally
of $O(1)$ according to the naive dimensional analysis \cite{georgi}.
Fig.~3a shows that the event rates/($100$\,${\rm fb}^{-1}$ GeV) from
operators $~{\cal L}_{4,5}~$ are larger than that from the $B$-term
when $~E=M_{WW} >600~$GeV, while the rates from operators
$~{\cal L}_{3,9,11;12}~$ can exceed $~|R_B|~$ only if $~E=M_{WW}>860$~GeV.
As $M_{WW}$ increases, the
rates contributed by $~{\cal L}_{4,5}~$ remain flat,
while the rates by $~{\cal L}_{3,9,11;12}~$  and the $B$-term decrease.
The ratio of the event rates from $~{\cal L}_{4,5}~$
to $~|R_B|~$ is $~5.0$ at $~E=M_{WW}=1$\,TeV, and rapidly increases
to $~19.6$ at $~E=M_{WW}=1.5~$\,TeV.
In contrast, the ratio between the rates from
$~{\cal L}_{3,9,11;12}~$  and the $B$-term
only varies from $~1.4$ to $3.0$ for $~E=M_{WW}=1\sim 1.5$\,TeV,  which
means that they are of the same order.
Fig.~3b shows that for the coefficients of $~O(1)~$, the event rates
contributed by operators $~{\cal L}^{(2)\prime}~$ and
$~{\cal L}_{1,2,8;13,14}~$  are all below $~|R_B|~$ for a wide
region of energy up to about $~2$\,TeV,
so that they cannot be sensitively probed in this case.
Especially, the contributions from
$~{\cal L}_{1,13}~$  are about two orders of magnitude
lower than that from the $B$-term.
This suggests that $~{\cal L}_{1,13}~$   must be tested via
other processes.
In Figs.~4a and 4b, different event rates are compared for
the coefficients ( $\ell_i$'s ) to be of $O(10)$.
Fig.~4a shows that
the rates from $~{\cal L}_{3,9,11;12}~$ could significantly dominate
$~|R_B|~$ by an order of magnitude for
$E=M_{WW}\sim 1$\,TeV
if their coefficients are increased by a factor
of $~10$ relative to the natural size of $~O(1)~$.
Fig.~4b shows that the rates from
$~{\cal L}_{1,13}~$ is still lower than $~|R_B|~$ by about an order of
magnitude,
while the rate from $~{\cal L}_{2}~$ is close to $~|R_B|~$  within
a factor of $~2$. The contributions from $~{\cal L}_{8;14}~$ and
$~{\cal L}^{(2)\prime}~$ exceed $~|R_B|~$ by about a factor
$2\sim 3$ at $~E=M_{WW}=1$\,TeV and a factor $3\sim 5$ at
$~E=M_{WW}=1.5$\,TeV when their coefficients are of $~O(10)~$.

{}From the above analyses, we conclude that studying the
$~W^+W^+\rightarrow W^+W^+~$ process can sensitively probe
the operators $~{\cal L}_{4,5}~$, but is only marginally sensitive for
probing $~{\cal L}_{3,9,11;12}~$ and insensitive for
$~{\cal L}^{(2)\prime}~$ and $~{\cal L}_{1,2,8;13,14}~$, if their
coefficients are naturally of $~O(1)~$. In the extreme case where
their coefficients are of $~O(10)~$, the probe of $~{\cal L}_{3,9,11;12}~$
could become sensitive and that of $~{\cal L}^{(2)\prime}~$ and
$~{\cal L}_{8;14}~$ could become marginally sensitive, while
$~{\cal L}_{2}~$  and $~{\cal L}_{1;13}~$ still cannot be sensitively
measured.

Finally, we note that the operators $~{\cal L}_{6,7,10}~$,
which violate the custodial $SU(2)_C$ symmetry,
do not contribute to the $W^+W^+$ pair productions up
to the one-loop order.
They can however contribute to the other scattering channels such as
$~WZ\rightarrow WZ~$, $~WW\rightarrow ZZ~$,
$~ZZ\rightarrow WW~$ and $~ZZ\rightarrow ZZ~$,
cf. Table~3.\footnote{ $~\,{\cal L}_{10}$ only contributes
to $~ZZ\rightarrow ZZ~$ channel.}~
By our order of magnitude estimates,
we conclude that they will give the similar kind of contributions
to the $WZ$ or $ZZ$ channel as
$~{\cal L}_{4,5}~$ give to the $W^+W^+$ channel.
This is because all these operators
contain four covariant derivatives [cf. (3.6)].

Before concluding this section, we would like to comment on
the $W^-W^- \rightarrow W^-W^-$ production process.
At the LHC (a pp collider), in the TeV region,
the luminosity of $W^-W^-$ is
typically smaller than that of $W^+W^+$ by a factor of $3 \sim 5$.
This is because in the TeV region, where the fraction
of momentum ($x$) of proton carried by the quark which
emitting the initial state $W$-boson is large
(for $x=\frac{E}{\sqrt{S}}\sim 0.1$), the parton luminosity is
dominated by the valence quark contributions.
Since in the large-$x$ region, the probability of finding a down-type
valence quark in the proton is smaller than finding an up-type
valence quark,
the luminosity of $W^-W^-$ is smaller than that of $W^+W^+$.
However, as long as there are enough $W^-W^-$ pairs detected,
which requires a large integrated luminosity of the machine and a
high detection efficiency of the detector, a similar conclusion
on probing the effective operators for the $W^+W^+$
channel can also be drawn for this channel.
For $M_{WW} > 1.5 $\,{\rm TeV}, the $W^-W^-$ production rate becomes
about an order of magnitude smaller than the $W^+W^+$ rate for any given
operator. Thus, this process could not be sensitive to all these
operators in this very high energy region.

\vspace{0.5cm}
\noindent
{\bf 6. Conclusions}
\vspace{0.3cm}

In this work, based upon our recent study
on the intrinsic connection between
the longitudinal weak-boson scatterings and probing the EWSB sector, we
first formulate the physical content of the ET as a criterion for
discriminating physical processes
which are sensitive/insensitive to probing
the EWSB mechanism [cf. Eqs.~(2.3)$\sim$(2.5)].
Then, we develop a precise power counting rule (3.7) for the EWCL,
from a natural generalization of
Weinberg's counting method for low energy QCD interaction.

Armed with this powerful counting rule and using the ET as the physical
criterion for probing the EWSB sector, we further systematically classify
the sensitivities of various scattering processes to
the complete set of bosonic operators at the level of $S$-matrix
elements (cf. Tables~1$\sim$3).
The power counting hierarchy in (4.1) governs the
order of magnitude of all relevant scattering amplitudes.

In the EWCL formalism, the leading contribution from the LNI $B$-term
is found to be model-independent and contains only the SM gauge coupling
constant [cf. (2.5) and Fig.~5c in the Appendix].
All other parts in the $B$-terms are further suppressed
by a factor $\frac{M_W}{E}$ or
$\left(\frac{(E,~gf_{\pi})}{\Lambda}\right)^2$
relative to the
leading contribution given in (2.5), cf. Table~2. Thus, they are
negligibly small and  insensitive to probing the EWSB sector.
It is important to note that the model-independent leading
$B$-term (2.5) provides a very useful criterion
for discriminating among sensitive, marginally sensitive,
 and insensitive contributions
from the various new physics effective operators in (3.6).

Finally, based on the above power counting analysis
combined with the effective-$W$
approximation, we phenomenologically  probe the EWSB sector
at the LHC via the weak-boson scattering in the
same-charged channel: $~W^\pm W^\pm\rightarrow W^\pm W^\pm$.
Computed from this simple power counting analysis,
our numerical results for the LHC production rates coincide with
those explicit calculations performed in the literature
well within a factor of $2$ (cf. Fig.~1).
This indicates that our power counting analysis can provide
an elegant grasp of the overall semi-quantitative physical picture.
We perform the first complete,
semi-quantitative survey on the
sensitivities of all fifteen next-to-leading-order
$CP$-conserving and $CP$-violating effective operators at the LHC
via the $W^+W^+$ channel. The results are shown in Figs.~3$\sim$4.
We find that, for this channel, when the  coefficients $\ell_n$'s
are naturally of $O(1)$,
$~{\cal L}_{4,5}~$ are most sensitive,
$~{\cal L}_{3,9,11;12}~$ are marginally sensitive, and
$~{\cal L}^{(2)\prime}~$ and $~{\cal L}_{1,2,8;13,14}~$ are insensitive.
For the extreme case where
the coefficients are of $~O(10)~$,
then the probe of $~{\cal L}_{3,9,11;12}~$
could become sensitive and that of $~{\cal L}^{(2)\prime}~$ and
$~{\cal L}_{8;14}~$ could become marginally sensitive, while
$~{\cal L}_{2}~$  and $~{\cal L}_{1;13}~$ still cannot be sensitively
measured via this process so that they must be
measured via other processes ({\it e.g.}, $q \bar q \rightarrow VV$).
 Up to the next-to-leading order,
the $SU(2)_C$-violating operators $~{\cal L}_{6,7,10}~$
do not contribute to this process. They, however, can be probed
via the $WZ$ and $ZZ$ productions.

A similar conclusion holds for the $W^-W^-$ channel
except that the event rate is lower by about a factor
of $3 \sim 5$ in the TeV region because the quark
luminosity for producing a $W^-W^-$ pair is smaller
than that for a $W^+W^+$ pair in pp collisions.

\vspace{1.0cm}
\noindent
{\bf Acknowledgements}~~~
We thank Sally Dawson for discussion on the effective-$W$ approximation
used in Ref.~\cite{BDV}, and William A. Bardeen for useful suggestion.
H.J.H. is supported in part by the U.S. DOE under grant DEFG0592ER40709;
Y.P.K. is supported by the NSF of China
and the FRF of Tsinghua University;
C.P.Y. is supported in part by the NSF under grant PHY-9309902.


\vspace{1.2cm}
\noindent
{\bf  Appendix: Validity of the ET in some special kinematic regions}
\vspace{0.3cm}

Here we examine the validity of the ET in some special kinematic regions
and its physical implication in probing the EWSB, which often cause
confusion in the literature.
It is known that there are kinematic regions
in which the Mandelstam variables
{}~$t$~ or ~$u$~ is small or even vanishing
despite the fact that $~\sqrt{s}\gg M_W~$ for high energy scatterings.
Therefore, the amplitude that contains a~$t$- or $u$-channel diagram
with massless photon field can generate
a kinematic singularity when the scattering angle $\theta$
approaches to $0$ or $\pi$.
In the following, we study in such special kinematic regions
whether the $B$-term [cf. (2.1)] can be safely ignored to validate
the ET and its physical consequence to probing the EWSB sector.

For illustration, let us consider the tree level
$~W^+_LW^-_L\rightarrow W^+_LW^-_L~$ scattering
in the chiral Lagrangian formalism.
Generalization to loop orders is obvious since the kinematic problem
 analyzed here only concerns the one-particle-reducible (1PR) internal
$W$, $Z$ or photon line in the $t$-channel (or $u$-channel) diagram.
Both the tree level $~W^+_LW^-_L\rightarrow W^+_LW^-_L~$ and
$~\pi^+\pi^-\rightarrow \pi^+\pi^-~$
amplitudes in the chiral Lagrangian
formalism contain contact diagrams, $s$-channel $Z$-exchange and
photon-exchange diagrams, and $t$-channel $Z$-exchange and
photon-exchange diagrams.
In the C.M. frame, the two tree-level amplitudes
$~T[W_L]~$ and $~T[{\rm GB}]~$ are precisely:
$$
\begin{array}{l}
T[W_L]=
 ig^2\left[ -(1+\kappa )^2 \sin^2\theta
+ 2\kappa (1+\kappa )(3\cos\theta -1)
-{\rm c}_{\rm w}^2
   \displaystyle\frac{4\kappa (2\kappa +3)^2\cos\theta}{4\kappa +3
  -{\rm s}_{\rm w}^2{\rm c}_{\rm w}^{-2}}\right.\\[0.4cm]
  \left.  +{\rm c}_{\rm w}^2
   \displaystyle\frac{8\kappa (1+\kappa )(1-\cos\theta )(1+3\cos\theta )
   +2[(3+\cos\theta )\kappa +2][(1-\cos\theta )\kappa -\cos\theta ]^2}
   {2\kappa (1-\cos\theta )+{\rm c}_{\rm w}^{-2}} \right]\\[0.4cm]
 +ie^2 \left[ -\displaystyle\frac{\kappa (2\kappa +3)^2\cos\theta}
    {\kappa +1} +4(1+\kappa )(1+3\cos\theta )+
   \displaystyle\frac{[(3+\cos\theta )\kappa +2][(1-\cos\theta )\kappa
       -\cos\theta ]^2}{\kappa (1-\cos\theta )} \right]~~,
\end{array}
\eqno(A1a)             
$$  \\
$$
\begin{array}{l}
T[{\rm GB}] =
ig^2 \left[\displaystyle\frac{(1+\cos\theta )}{2}\kappa
           +\frac{1}{3}
+\displaystyle\frac{({\rm c}_{\rm w}^2-{\rm s}_{\rm w}^2)^2}
{2{\rm c}_{\rm w}^2}\left(-\displaystyle\frac{2\kappa\cos\theta}
{4\kappa +3-{\rm s}_{\rm w}^2{\rm c}_{\rm w}^{-2}}
+\displaystyle\frac{(3+\cos\theta )\kappa +2}{2(1-\cos\theta )\kappa
+ {\rm c}_{\rm w}^{-2}}\right)\right]\\[0.4cm]
+ie^2\left[ -\displaystyle\frac{4\kappa\cos\theta}{4\kappa +1}
+\displaystyle\frac{(3+\cos\theta )
   \kappa +2}{(1-\cos\theta )\kappa}\right]~~, \\[0.4cm]
\end{array}
\eqno(A1b)            
$$\\[0.10cm]
where $~\kappa \equiv p^2/M_W^2~$ with $~p~$ equal to the C.M. momentum;
$~{\rm s}_{\rm w}\equiv \sin\theta_{\rm W}~$,
      $~{\rm c}_{\rm w}\equiv \cos\theta_{\rm W}~$ with $\theta_{\rm W}$
equal to the weak mixing angle;
and $\theta$ is the scattering angle.
In (A1a) and (A1b) the terms without a momentum factor in the denominator
come from contact diagrams, terms with denominator independent of
scattering angle come from $s$-channel diagrams and terms with denominator
containing a factor $~1-\cos\theta~$
are contributed by $t$-channel diagrams.
Let us consider two special kinematic regions defined below.

\noindent
(i). In the limit of ~$\theta \rightarrow 0$:

As $~\theta\rightarrow 0~$, the $t$-channel photon propagator
has a kinematic pole, but
both $W_L$ and GB amplitudes have the {\it same} pole structure,
i.e.
$$
(T[W_L]-T[{\rm GB}])_{\rm pole~term}
=ie^2(2\kappa^{-1}+3+\cos\theta )[(1-\cos\theta )\kappa^2
-2 \kappa\cos\theta -(1+\cos\theta )]~~,
\eqno(A2)                      
$$
which is finite.\footnote{
This conclusion can be directly generalized to other $t$- or
$u$-channel processes.}
Hence, {\it the B-term,
which is defined as the difference $~T[W_L]-T[{\rm GB}]~$,
is finite at $~\theta =0~$, and is of
$~O(e^2)$ which is smaller than $O(g^2)$.}
This means that when $~\theta~$ is close to the $t$-channel photon pole,
the $B$-term is negligibly small relative to the GB-amplitude
so that (2.3b) is satisfied and
the ET works. More explicitly,  in the limit of
$~\theta =0~$ (i.e. $~t=0~$), and from (A1a,b),
the $W_L$ and GB amplitudes are
$$
\begin{array}{ll}
T[W_L] & = i\left[ 4(3-8{\rm c}_{\rm w}^2+8{\rm c}_{\rm w}^4)
\displaystyle\frac{p^2}{f_\pi^2}
+2e^2\left( 2+\displaystyle\frac{M_W^2}{p^2}\right)\frac{1}{1-c_0}\right]
 +O(g^2)~~,\\[0.45cm]
T[{\rm GB}] & = i\left[ 4(3-8{\rm c}_{\rm w}^2+8{\rm c}_{\rm w}^4)
\displaystyle\frac{p^2}{f_\pi^2}
+2e^2\left( 2+\displaystyle\frac{M_W^2}{p^2}\right)
\displaystyle\frac{1}{1-c_0}\right]
 +O(g^2)~~,\\[0.4cm]
T[W_L] & = T[{\rm GB}] + O(g^2) ~~,
\end{array}
\eqno(A3)                              
$$
where $~~c_0\equiv \lim_{\theta\rightarrow 0}\cos\theta~$.~
In this case one cannot make the $M_W^2/t~$ expansion\footnote{This
expansion is {\it unnecessary} for the validity of the ET,
cf. (2.3) and (2.3a,b).}~~ because ~$t$ vanishes identically. Since
both $W_L$ and GB amplitudes have exactly the same kinematic
singularity and the $B$-term is much smaller than $T[{\rm GB}]$,
{\it the ET still holds} in this special kinematic region.
We also emphasize that {\it in the
kinematic regions where $t$ or $u$ is not much larger than $M_W^2$,
the $t$-channel or $u$-channel
internal gauge boson lines must be included
according to the precise formulation of the ET}
[cf. (2.3) and (2.3a,b)].\footnote{
This does not imply, in any sense, a violation of the ET
since the ET, cf. (2.3) and (2.3a,b), does not require
either $t \gg M_W$ or $u \gg M_W$.}

\noindent
(ii). In the limit of ~$\theta \rightarrow \pi$:

In this kinematic region, $~s,~t\gg M_W^2~$, and (A1a,b) yield
$$
\begin{array}{ll}
T[W_L] & =i\displaystyle\left[ 2(1+\cos\theta )\frac{p^2}{f_\pi^2}
                   +O(g^2)\right] ~~,\\[0.4cm]
T[{\rm GB}] & = i\displaystyle\left[ 2(1+\cos\theta )\frac{p^2}{f_\pi^2}
                   +O(g^2)\right]  ~~,\\[0.55cm]
T[W_L] & = T[{\rm GB}] + O(g^2) ~~,
\end{array}                                     
\eqno(A4)
$$
where the $~O(g^2)~$ term is the largest term
we ignored which denotes the
order of the $B$-term [cf. (2.5)];
all other terms we ignored such as
$~O(M_W^2/p^2)~$ or $~O(e^2)~$
are smaller than $O(g^2)$  and thus will not affect
the order of magnitude estimate of the $B$-term.
For $~s,~t\gg M_W^2~$, the $W_L$ and GB amplitudes are dominated
by the $p^2$-term in (A4), which is actually proportional to $u$ for
this process.
When the scattering angle $\theta$ is close to $180^{\circ}$,
$u$ becomes small and
thus this leading $p^2$ term is largely suppressed
so that both the $W_L$ and GB amplitudes can be
as small as the $B$-term, i.e.
of  $O(g^2)$. In this case our condition (2.3a) is satisfied
while (2.3b) is not, which means that the EWSB sector
cannot be sensitively probed for this kinematic region. Since
the total cross section of this process
is not dominated by this special kinematic region and
is mainly determined by the un-suppressed
leading large $p^2$-term,
so {\it the kinematic dependence of the amplitude
will not affect the order of magnitude of the total cross section}.
Hence, {\it our application of the power counting analysis in Sec.~5 for
computing the total event rates remains valid even
though we have ignored the angular dependence in
estimating the magnitude of the scattering amplitudes.}
Neglecting the angular dependence in the amplitude may
cause a small difference in the event rate
as compared to that from detailed precise calculations.
For the processes such as
$~W^\pm_L W^\pm_L\rightarrow W^\pm_L W^\pm_L ~$ and
$~W_L^+ W^-_L \rightarrow Z_LZ_L~$, the leading $p^2$-term
is proportional to $~s/f_\pi^2~$ with no angular dependence, so that
the angular integration causes {\it no difference} between our power
counting analysis and the exact calculation for the leading
$p^2$-term contribution.\footnote{
The small difference (a factor of 1.4) in Fig.~1
mainly comes from  neglecting  the tree level sub-leading terms
in our order of magnitude estimate for the amplitudes.}~
In the above example for $~W^+_L W^-_L\rightarrow W^+_L W^-_L ~$ channel
[cf. (A4)], the leading amplitude is proportional to $~-u/f_\pi^2~$.
Using the power counting method,
we ignore the $\theta$-dependence and estimate it as $~s/f_\pi^2~$.
In computing the total rate, we integrate out the scattering angle.
This generates a difference from the precise one:
$$
{ {\int_{-1}^{1} u^2 \, {\rm d}\! \cos \theta} \over
{\int_{-1}^{1} s^2 \,   {\rm d}\! \cos \theta} } = \frac{1}{3} ~~,
$$
which, as expected, is only a factor of $3$ and does not affect our order
of magnitude estimates.

Finally, we make a precise numerical analysis on the equivalence between
the $W_L$ and GB amplitudes to show how well the ET works in different
kinematic regions and its implication to probing the EWSB sector.
We use the full expressions (A1a,b) for $W_L$ and GB amplitudes
as required by the ET, cf. (2.3) and (2.3a,b).
In Fig.~5a, we plot the ratio $~|B/g^2|~$ for scattering angle
$~\theta = 2^{\circ}, 10^{\circ}, 45^{\circ},
90^{\circ}, 100^{\circ}, 120^{\circ}, 135^{\circ},
150^{\circ}, 180^{\circ}~$.
Fig.~5a shows that the LNI $B$-term is {\it always of $~O(g^2)~$
in the whole kinematic region,} and thus is irrelevant to
the EWSB sector,  in accord with our general physical analysis in Sec.~2.
Hence, to have a sensitive probe of the EWSB mechanism, condition
(2.3b) or (2.4) must be satisfied.
Fig.~5b shows that
for $~0\leq \theta \leq 100^{\circ}~$,
the ratio $~~|B/T[W_L]|\leq 10\% ~~$ when
$~M_{WW}\geq 500$\,GeV.
For $~\theta \geq 120^{\circ}~$, this ratio becomes
large and reaches $~O(1)~$ when $~\theta~$ is close to $180^{\circ}$.
This is because the kinematic factor
$(1+\cos\theta )$, associated with the leading $p^2$ term [cf. (A4)],
becomes small. This, however, will
not alter the conclusion that for $4W_L$-scattering the total cross
section from $T[{\rm GB}]$ is much larger than that from the $B$-term
as $~M_{WW}\geq 500$\,GeV.
Note that in Fig.~5b, for  $~\theta \leq 10^{\circ}~$, i.e. close to the
$t$-channel photon pole,
 the ratio $~|B/T[W_L]|~$ is below $~1\% ~$ and thus the ET holds
very well.
In Fig.~5c, we plot both the $W_L$ and GB amplitudes for
$~\theta = 10^{\circ},45^{\circ}, 100^{\circ},
150^{\circ}~$. The solid lines denote the
complete $W_L$ amplitude and the other lines denote the GB amplitude.
We find that when $~\theta \leq 100^{\circ}~$, the GB amplitude is almost
indistinguishable from the $W_L$ amplitude. For $~\theta =150^{\circ}~$,
the $W_L$ amplitude is of the same order as the $B$-term, i.e. of
$~O(g^2)~$, when $~M_{WW}< 1$\,TeV. In this case the $W_L$ or GB
amplitude is
too small and the strongly coupled EWSB sector cannot be sensitively
probed. As the energy $E$ increases, we see that the $W_L$ and GB
amplitudes rapidly dominate over the $B$-term and agree
better and better even for large scattering angles.

The above conclusions hold for the tree level contributions from
the lowest order operators in
$~{\cal L}_G +{\cal L}^{(2)}+{\cal L}_F~$, cf. (3.6).
However, independent of the kinematic region considered,
not all the contributions from the
next-to-leading-order effective operators
can dominate the $B$-term and satisfy the condition (2.3b).
This is why the condition (2.3b) can serve as the
criterion for classifying the sensitivities of these
next-to-leading-order operators
in probing the EWSB sector for each given process.

We conclude that {\it the B-term as
defined in (2.1) can be at most of $O(g^2)$
for all kinematic regions} (cf. Fig.~5a),
and is insensitive to the EWSB
mechanism, in accord with our general analysis in Sec.~2.
{\it When ~$t$ or ~$u$
is not large, the ~$t$- or ~$u$-channel
internal  lines must be included}.
We find that {\it even for $t$ or $u$ being close to zero,
the ET still works well} [cf. Eq.~(A3) and Fig.~5b].
This is because
the validity of the ET does not require either
$~t \gg M_W^2$ or $~u \gg M_W^2$, cf. (2.3) and (2.3a,b).
 For some scattering processes,
there may be special kinematic regions in which the GB and $W_L$
amplitudes are largely suppressed\footnote{
This large suppression can also arise from the polarization effects
of the in/out states.}~
 so that the EWSB sector cannot be sensitively  probed
in these special kinematic regions.
However, it can still be sensitively probed
by measuring the total event rates from these processes.

\newpage

\newpage
\noindent
{\bf Table Captions}
\vspace{0.2cm}

\noindent
{\bf Table 1.}  Estimates of amplitudes
for $~W^\pm W^\pm \rightarrow W^\pm W^\pm ~$
scattering.\\[0.35cm]
{\bf Table 1a.} Model-independent contributions from
$~{\cal L}_G +{\cal L}_F +{\cal L}^{(2)}~$.\\[0.20cm]
{\bf Table 1b.} Model-dependent contributions from the
next-to-leading-order
effective operators.

\noindent
{\bf Table 2.} Order estimates of $B$-terms for
$~W^\pm W^\pm \rightarrow W^\pm W^\pm ~$ scattering.\\[0.35cm]
{\bf Table 2a.}  Model-independent contributions.\\[0.20cm]
{\bf Table 2b.}  Relevant operators for model-dependent
                 contributions.$^{(a)}$ \\
Notes:\\ {\footnotesize
$^{(a)}$ We list the relevant operators for each order of $B$-terms. \\
$^{(b)}$ Here $B^{(0)}_1$ is contributed by $T_1[2\pi^\pm ,2v^\pm ]$.}

\noindent
{\bf Table 3.} Global classification of sensitivities
to probing direct and
indirect EWSB information
from effective operators at the level of $S$-matrix
elements.$^{(a)}$\\
Notes:\\ {\footnotesize
$^{(a)}$ The contributions from
${\cal L}_{1,2,13}$ are {\it always} associated
with a factor of $\sin^2\theta_W$,
unless specified otherwise.\\
$^{(b)}$ MI $=$ model-independent, MD $=$ model-dependent.\\
$^{(c)}$ There is no contribution when all the external lines are
electrically neutral.\\
$^{(d)}$ $B_0^{(1)}\simeq T_0[2\pi ,v,V_T]~(\neq T_0[2\pi^0 ,v^0,Z_T])$,~
$B_0^{(3)}\simeq T_0[v,3V_T]~(\neq T_0[v^0,3Z_T])$.\\
$^{(e)}$  $T_1[2V_L,2V_T]=T_1[2Z_L,2W_T],~T_1[2W_L,2Z_T]$, ~or
$~T_1[Z_L,W_L,Z_T,W_T]$.\\
$^{(f)}$ ${\cal L}_2$ only contributes to $T_1[2\pi^\pm ,\pi^0,v^0]$ and
$T_1[2\pi^0,\pi^\pm ,v^\pm ]$ at this order;
${\cal L}_{6,7}$ do not contribute
to $T_1[3\pi^\pm ,v^\pm ]$.\\
$^{(g)}$  ${\cal L}_{10}$ contributes only
to $T_1[\cdots ]$ with all the external
lines being electrically neutral.\\
$^{(h)}$ Here, $T_1[2W_L,2W_T]$ contains a coupling
$e^4=g^4\sin^4\theta_W$.\\
$^{(i)}$ ${\cal L}_2$ only contributes to $T_1[3\pi^\pm ,v^\pm ]$.\\
$^{(j)}$ ${\cal L}_{1,13}$ do not contribute to
$T_1[2\pi^\pm ,2v^\pm ]$.  }

\newpage
\noindent
{\bf Figure Captions}
\vspace{0.2cm}

\noindent
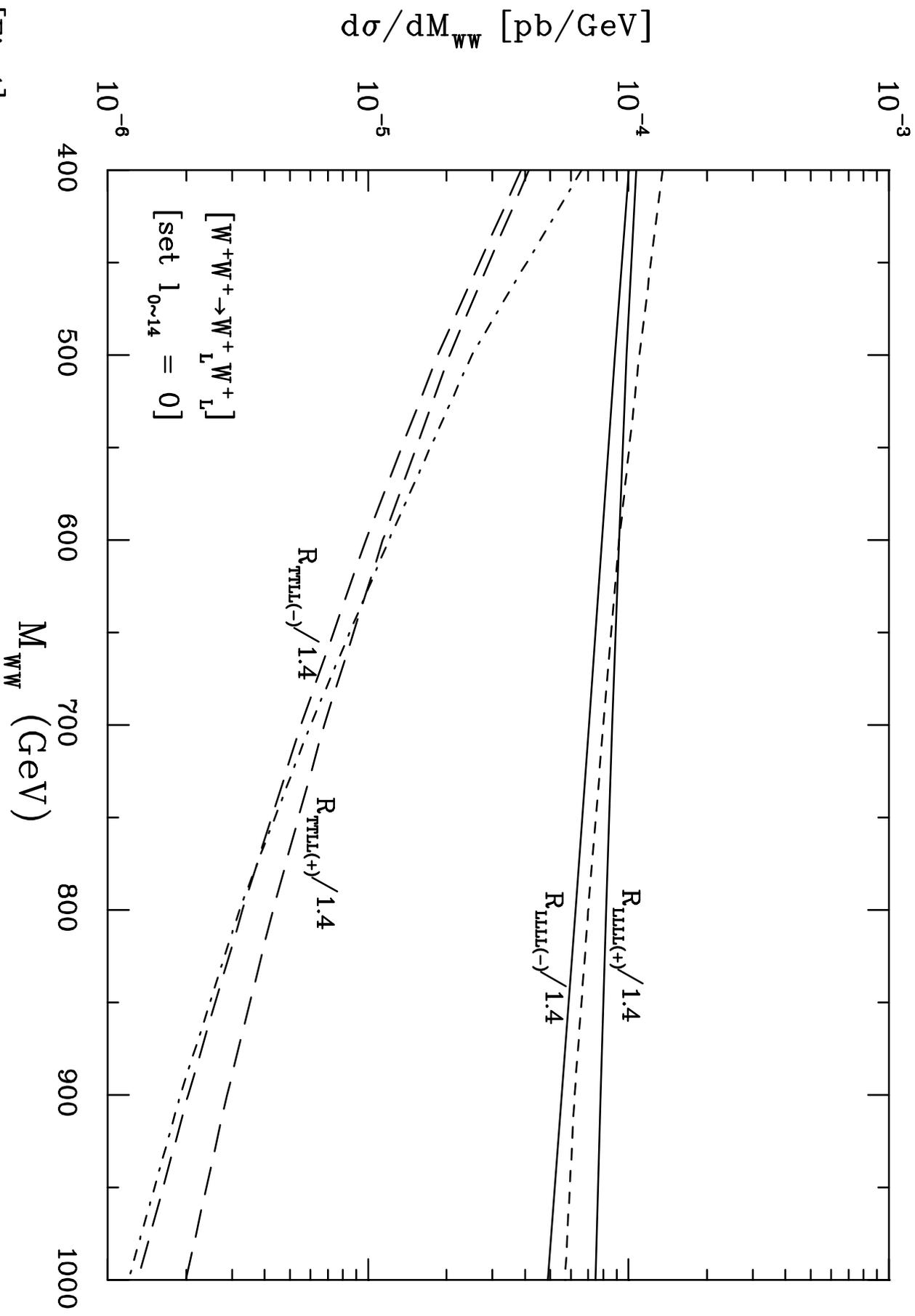
{\bf Fig.~1.}  Comparison with the Fig.~8 of Ref.~\cite{BDV}
up to $1$-loop for
$\sqrt{S}= 40$\,TeV.
The solid and long-dashed lines are given by our power counting analysis.
The dashed and dot-dashed lines are $R_{LLLL}$ and $R_{TTLL}$ of
Ref.~\cite{BDV}
which coincide with ours within a factor of 2.
[The meanings of the production rates
$R_{\alpha\beta\gamma\delta}$'s are defined in the text,
cf. (5.1a,b).]

\noindent
{\bf Fig.~2.}\\
{\bf (2a).} Comparison of the production rates of $W^+_LW^+_L$ pairs
up to $1$-loop for the
$W^+_LW^+_L$, $W^+_LW^+_T$ and $W^+_TW^+_T$ initial states,
at the  $14$\,TeV LHC.\\
{\bf (2b).} Comparison of the production rates of
different final states up to $1$-loop
after summing over the
polarizations of the initial states, at the $14$\,TeV LHC.

\noindent
{\bf Fig.~3.} Sensitivities of operators
${\cal L}^{(2)\prime}$ and ${\cal L}_{1\sim 14}$,
when their coefficients are of
$O(1)$, at the $14$ TeV LHC.\\[0.20cm]
{\bf (3a).} For operators ${\cal L}_{4,5,3,9,11,12}~$.\\
{\bf (3b).} For operators
${\cal L}^{(2)\prime}$ and ${\cal L}_{1,2,8,13,14}~$.

\noindent
{\bf Fig.~4.}  Same as Fig.~3, but the coefficients $\ell_n$'s are
of $O(10)$.\\[0.20cm]
{\bf (4a).} For operators ${\cal L}_{4,5,3,9,11,12}~$.\\
{\bf (4b).} For operators
${\cal L}^{(2)\prime}$ and ${\cal L}_{1,2,8,13,14}~$.

\noindent
{\bf Fig.~5.} Examination on the kinematic dependence and the
validity of the ET
for the $~W^+_LW^-_L \rightarrow W^+_LW^-_L ~$ scattering
process.   \\[0.20cm]
{\bf (5a).}  The ratio $~|B/g^2|~$ for $~\theta
= 2^{\circ},~10^{\circ},~45^{\circ},~
90^{\circ},~100^{\circ},~120^{\circ},~ 135^{\circ},~ 150^{\circ},~
180^{\circ}~$. \\
{\bf (5b).}  Same as (5a), but for the ratio $~|B/T[W_L]|~$.\\
{\bf (5c).}  Comparison of the $W_L$-amplitude (solid lines) and
the corresponding GB-amplitude (non-solid lines) for
$~\theta = 10^{\circ},~45^{\circ},~100^{\circ},~150^{\circ}~$.
(Here, $~B[150^{\circ}]~$ denotes
the $B$-term at $~\theta =150^{\circ}~$.)


\addtolength{\textwidth}{2.0cm}
\addtolength{\oddsidemargin}{+0.7cm}
\evensidemargin=\oddsidemargin
\addtolength{\textheight}{1.0cm}
\addtolength{\topmargin}{-0.8cm}

\newpage

\renewcommand{\baselinestretch}{1}


\begin{table}[t]
\begin{center}

{\bf Table 1.}  Estimates of
amplitudes for $~W^{\pm}W^{\pm}\rightarrow W^{\pm}W^{\pm}~$ scattering.
\vspace{0.8cm}

{\bf Table 1a.}
{}~Model-independent contributions from
$~{\cal L}_G+{\cal L}_F +{\cal L}^{(2)}~$.
\vspace{0.5cm}


\small

\begin{tabular}{||c||c|c|c|c|c||}
\hline\hline
& & & & &  \\
${\cal L}_G +{\cal L}_F + {\cal L}^{(2)}$
&  $~~~T_{\ell}[4\pi]~~~     $
&  $~T_{\ell}[3\pi,W_T]~ $
&  $~T_{\ell}[2\pi,2W_T]~$
&  $~T_{\ell}[\pi,3W_T]~ $
&  $~T_{\ell}[4W_T]~     $  \\
& & & & &  \\
\hline\hline
& & & & &\\
Tree-Level
&  $ \frac{E^2}{f_{\pi}^2} $
&  $ g\frac{E}{f_\pi} $
&  $ g^2 $
&  $ e^2g\frac{f_\pi}{E} $
&  $g^2$ \\
( ${\ell}=0$ ) & & & & &\\
& & & & &\\
\hline
& & & & &\\
One-Loop
& $\frac{E^2}{f_{\pi}^2}\frac{E^2}{\Lambda^2}$
& $g\frac{E}{f_{\pi}}\frac{E^2}{\Lambda^2}$
& $g^2\frac{E^2}{\Lambda^2}$
& $g^3\frac{f_{\pi}E}{\Lambda^2}$
& $g^4\frac{f_{\pi}^2}{\Lambda^2}$   \\
( ${\ell}=1$ ) & & & & &\\
& & & & & \\
\hline\hline
\end{tabular}
\end{center}
\end{table}

\vspace{0.8cm}


\newpage

\begin{table}[t]
\begin{center}
{\bf Table 1b.} Model-dependent contributions from the
next-to-leading-order  effective operators.
\end{center}
\vspace{20.00cm}
\end{table}


\newpage

\renewcommand{\baselinestretch}{1}

\begin{table}[t]
\begin{center}

{\bf Table 2.} Order estimates of $B$-terms
for $~W^{\pm}W^{\pm}\rightarrow W^{\pm}W^{\pm}~$  scattering.
\vspace{0.8cm}

{\bf Table 2a.} Model-independent contributions.
\vspace{0.5cm}


\small

\begin{tabular}{||c||c|c|c|c||}
\hline\hline
& & & &  \\
${\cal L}_G +{\cal L}_F + {\cal L}^{(2)}$
&  $~~~B_{\ell}^{(0)}~~~     $
&  $~~~B_{\ell}^{(1)}~~~     $
&  $~~~B_{\ell}^{(2)}~~~     $
&  $~~~B_{\ell}^{(3)}~~~     $  \\
& & & &   \\
\hline\hline
& & & & \\
Tree-Level
&  $ g^2$
&  $ g^2\frac{M_W}{E} $
&  $ e^2\frac{M_W^2}{E^2} $
&  $ g^2\frac{M_W}{E} $\\
 ( ${\ell}=0$ ) & & & &\\
& & & & \\
\hline
& & & & \\
One-Loop
& $g^2\frac{E^2}{\Lambda^2}$
& $g^3\frac{Ef_\pi}{\Lambda^2}$
& $ g^4\frac{f_{\pi}^2}{\Lambda^2}$
& $g^4\frac{f_{\pi}^2}{\Lambda^2}\frac{M_W}{E} $\\
( ${\ell}=1$ ) & & & &\\
& & & &  \\
\hline\hline
\end{tabular}

\vspace{1.5cm}


{\bf Table 2b.} Relevant operators for
model-dependent contributions.$^{(a)}$
\vspace{0.8cm}


\small

\begin{tabular}{||c|c|c|c||}
\hline\hline
& & &   \\
    $O(g^2\frac{E^2}{\Lambda^2})$
&    $O(g^3\frac{Ef_\pi}{\Lambda^2})$
&    $O(g^2\frac{f_{\pi}^2}{\Lambda^2})$
&    $O(g^4\frac{f_{\pi}^2}{\Lambda^2})$ \\
    ( from $B^{(0)}_1$ )
&   ( from $B^{(1)}_1$ )
&   ( from $B^{(0)}_1$ )
&   ( from $B^{(2)}_1 ~{\rm or} ~B^{(0)}_1 $ )\\
 & & &   \\
\hline
 & & & \\
  ${\cal L}_{3,4,5,9,11,12}$
& ${\cal L}_{2,3,4,5,8,9,11,12,14}$
& ${\cal L}^{(2)\prime}$
& \parbox[t]{4.28cm}{
${\cal L}_{1\sim 5,8,9,11\sim 14} ~~(B^{(2)}_1)$ \\
${\cal L}_{1,2,8,13,14} ~~(B^{(0)}_1)$ \\
${\cal L}_{2\sim 5,8,9,11,12,14} ~(B^{(0)}_1)~^{(b)}$ } \\
& & & \\
\hline\hline
\end{tabular}
\end{center}
\end{table}




\addtolength{\textheight}{0.5cm}
\addtolength{\topmargin}{-0.5cm}
\addtolength{\oddsidemargin}{-0.6cm}
\addtolength{\oddsidemargin}{-0.8cm}
\evensidemargin=\oddsidemargin
\newpage


\begin{table}[t]
\begin{center}

{\bf Table 3.}
Global classification of sensitivities to probing direct and indirect EWSB\\
information from effective operators at the level of $S$-matrix elements.
$^{(a)}$

\vspace{0.5cm}


\small

\begin{tabular}{||c||c|c|c||}
\hline\hline
& & &  \\
Required Precision
&  Relevant Operators
&  Relevant Amplitudes
&  MI or MD $^{(b)}$ \\
& & & ? \\
\hline\hline
   $O(\frac{E^2}{f_{\pi}^2})$
&  ${\cal L}_0~(\equiv {\cal L}_G+{\cal L}_F+{\cal L}^{(2)}) $
&  $ T_0[4V_L] (\neq T_0[4Z_L]) $
&  MI \\
\hline
\parbox[t]{2.2cm}{
{}~~\\
{}~~\\
$O(\frac{E^2}{f_\pi^2}\frac{E^2}{\Lambda^2},~g\frac{E}{f_\pi}$)\\
{}~~\\
{}~~ }
&  \parbox[t]{2.8cm}{
  ${\cal L}_{4,5}$\\
  ${\cal L}_{6,7}$\\
  ${\cal L}_{10}$\\
  ${\cal L}_0$\\
  ${\cal L}_0$ }
&  \parbox[t]{5.0cm}{
  $T_1[4V_L]$\\
  $T_1[2Z_L,2W_L],~T_1[4Z_L]$\\
  $T_1[4Z_L]$\\
  $T_0[3V_L,V_T] ~(\neq T_0[3Z_L,Z_T])$\\
  $T_1[4V_L]$\\ }
&  \parbox[t]{0.8cm}{
  MD\\
  MD\\
  MD\\
  MI\\
  MI }\\
\hline
\parbox[t]{2.2cm}{
{}~~\\
{}~~\\
$O(g\frac{E}{f_\pi}\frac{E^2}{\Lambda^2},~g^2)$\\
{}~~\\
{}~~\\
{}~~  }
& \parbox[t]{2.8cm}{
  $ {\cal L}_{3,4,5,9,11,12} $\\
  $ {\cal L}_{2,3,4,5,6,7,9,11,12} $\\
  $ {\cal L}_{3,4,5,6,7,10} $\\
  $ {\cal L}_0 $\\
  $ {\cal L}_0 $\\
  $ {\cal L}_0 $}
&  \parbox[t]{5.8cm}{
  $T_1[3W_L,W_T]$\\
  $T_1[2W_L,Z_L,Z_T],~T_1[2Z_L,W_L,W_T]$\\
  $T_1[3Z_L,Z_T]$\\
  $T_0[2V_L,2V_T],~T_0[4V_T]~~^{(c)}$\\
  $T_1[3V_L,V_T]$\\
  $B^{(0)}_0 \simeq T_0[3\pi ,v]~(\neq T_0[3\pi^0 ,v^0] )$  }
&   \parbox[t]{0.8cm}{
  MD\\
  MD\\
  MD\\
  MI\\
  MI\\
  MI }\\
\hline
  $O(\frac{E^2}{\Lambda^2})$
& ${\cal L}^{(2)\prime}$
& $T_1[4W_L],~T_1[2W_L,2Z_L]$
& MD \\
\hline
   \parbox[t]{2.5cm}{
        ~~\\
{}~~\\
{}~~\\
$O(g^2\frac{E^2}{\Lambda^2},~g^3\frac{f_\pi}{E})$ \\
{}~~\\
{}~~\\
{}~~ }
&  \parbox[t]{3.0cm}{
   ${\cal L}_0$\\
   ${\cal L}_{2,3}$\\
   ${\cal L}_{3,11,12}$\\
   ${\cal L}_{2,3,4,5,8,9,11,12,14}$\\
   ${\cal L}_{1\sim 9,11\sim 14}$ \\
   ${\cal L}_{4,5,6,7,10}$ \\
   ${\cal L}_{0,2,3,4,5,6,7,9\sim 12}$ }
&  \parbox[t]{6.3cm}{
   $T_0[V_L,3V_T],T_1[2V_L,2V_T], B^{(1,3)}_0~~^{(c,d)} $\\
   $T_1[4W_L]$\\
   $T_1[2Z_L,2W_L]$\\
   $T_1[2W_L,2W_T]$\\
   $T_1[2V_L,2V_T]~~~^{(e)}$\\
   $T_1[2Z_L,2Z_T]$\\
   $B^{(0)}_1 \simeq T_1[3\pi ,v]~~^{(f,g)}$ }
&  \parbox[t]{2.0cm}{
  MI\\
  MD\\
  MD\\
  MD\\
  MD\\
  MD\\
  MI $+$ MD\\  }\\
\hline
\parbox[t]{1.75cm}{
{}~~\\
{}~~\\
  $O(g^3\frac{Ef_\pi}{\Lambda^2})$\\
{}~~\\}
& \parbox[t]{3.3cm}{
${\cal L}_{0,1,2,3,8,9,11\sim 14}$ \\
${\cal L}_{4,5}$\\
${\cal L}_{6,7,10}$ \\
${\cal L}_{2\sim 5,8,9,11,12,14}$ }
& \parbox[t]{5.5cm}{
$T_1[V_L,3V_T]~(\neq T_1[Z_L,3Z_T])$\\
$T_1[V_L,3V_T]$\\
$T_1[V_L,3V_T]~(\neq T_1[W_L,3W_T])~~^{(g)}$ \\
$B^{(1)}_1\simeq T_1[2\pi ,V_T,v]$  }
& \parbox[t]{2.0cm}{ MI$+$MD \\
                     MD\\
                     MD\\
                     MD }\\
\hline
\parbox[t]{2.0cm}{
{}~~\\
{}~~\\
{}~~\\
{}~~\\
{}~~\\
  $O((g^2,g^4)\frac{f_\pi^2}{\Lambda^2})$\\
{}~~\\
{}~~\\
{}~~\\
{}~~\\
{}~~}
& \parbox[t]{3.1cm}{
  ${\cal L}^{(2)\prime} $ \\
  ${\cal L}_1$\\
  ${\cal L}_{0,1\sim 5,8,9,11\sim 14}$\\
  ${\cal L}_{0,1\sim 9,11\sim 14}$\\
  ${\cal L}_{0,1,4,5,6,7,10}$\\
  ${\cal L}_{1,2,8,13,14}$\\
  ${\cal L}_{0,1\sim 9,11\sim 14}$\\
  ${\cal L}_{0,4,5,6,7,10}$\\
  ${\cal L}_{0,1\sim 5,8,9,11\sim 14}$\\
  ${\cal L}_{0,1\sim 9,11\sim 14}$\\
  ${\cal L}_{0,4,5,6,7,10}$ }
&
\parbox[t]{6.4cm}{
$T_1[2V_L,2V_T],B^{(0)}_1\simeq T_1[3\pi ,v]~~^{(c)}$ \\
  $ T_1[2W_L,2W_T]~~^{(h)} $\\
  $ T_1[4W_T] $\\
  $ T_1[4V_T]~(\neq T_1[4W_T],T_1[4Z_T]) $\\
  $ T_1[4Z_T] $\\
{\footnotesize  $ B_1^{(0)}\simeq T_1[3\pi ,v]~~^{(c,i)} $\\
  $ B_1^{(0)}\simeq T_1[2\pi ,2v]~~^{(c,j)} $\\
  $ B_1^{(0)}\simeq
                      T_1[2\pi ,2v](\neq T_1[2\pi^\pm ,2v^\pm ])~^{(g)} $ \\
  $ B_1^{(2)}\simeq T_1[\pi^\pm ,2W_T,v^\pm ] $\\
  $ B_1^{(2)}\neq T_1[\pi^\pm ,2W_T,v^\pm ],
     T_1[\pi^0 ,2Z_T,v^0] $\\
  $ B_1^{(2)}\simeq T_1[\pi^0 ,2Z_T,v^0] $ } }
&  \parbox[t]{2.0cm}{
  MD\\
  MD\\
  MI$+$MD\\
  MI$+$MD\\
  MI$+$MD\\
  MD\\
  MI$+$MD\\
  MI$+$MD\\
  MI$+$MD\\
  MI$+$MD\\
  MI$+$MD  }\\
& & &  \\
\hline\hline
\end{tabular}
\end{center}
\end{table}



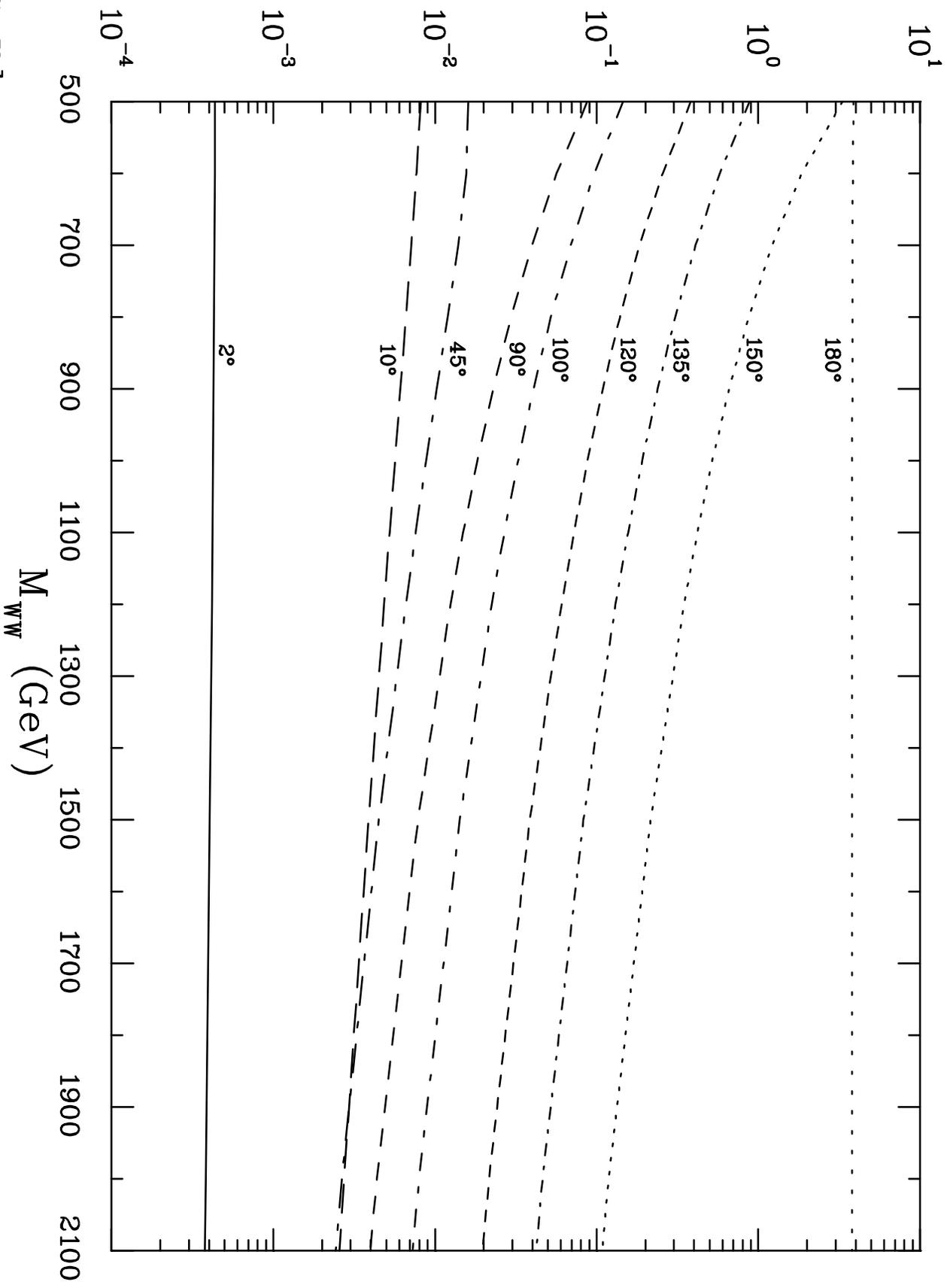
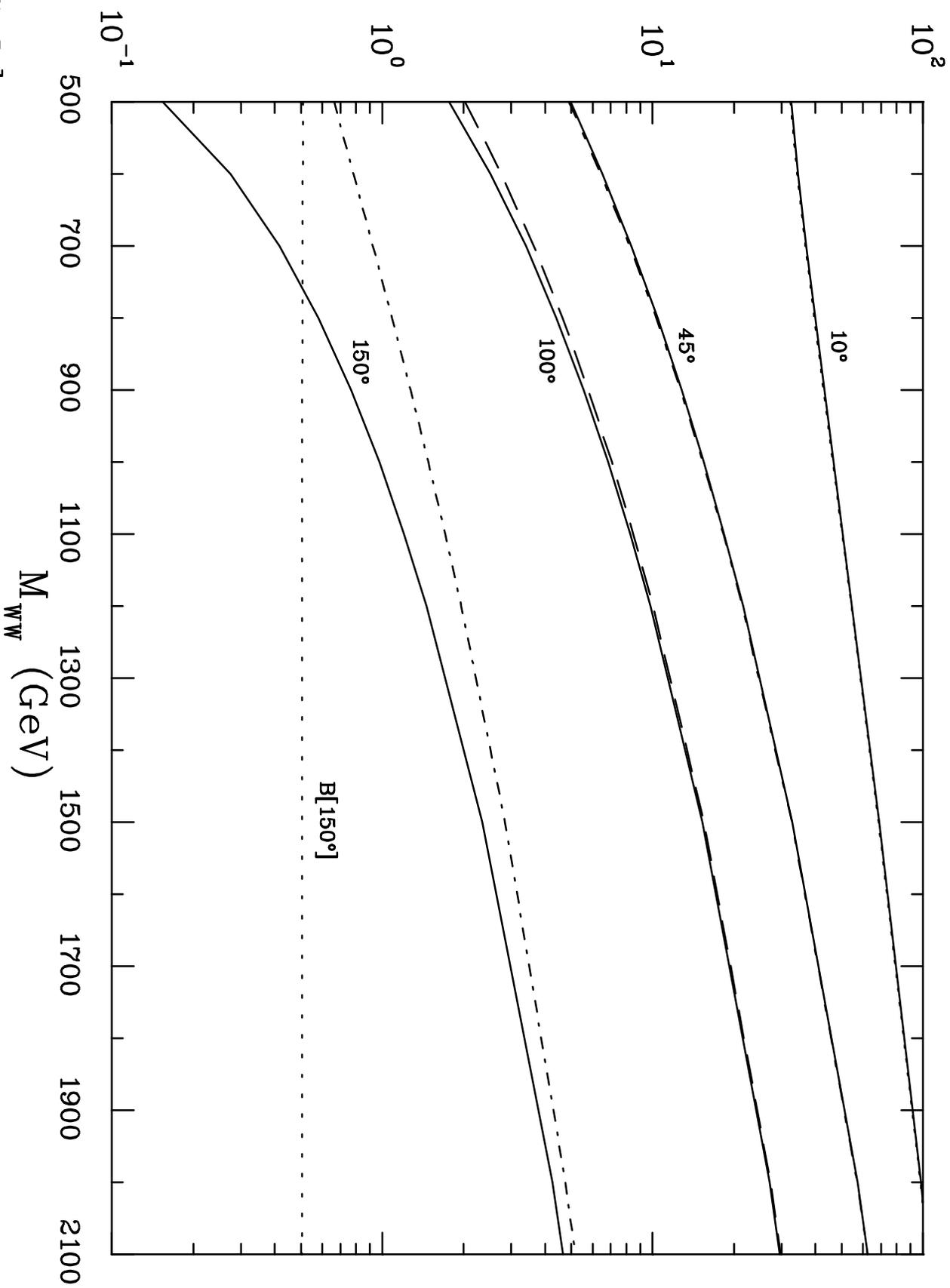
\end{document}